\documentclass{article}

\usepackage{graphicx}
\usepackage{natbib}
\usepackage{color}
\def\aaa{{\it a}}
\def\bbb{{\it b}}
\def\ccc{{\it c}}
\newcommand{\mylab}[3]{\raisebox{#2}[0mm][0mm]{%
\makebox[0mm][l]{\hspace*{#1}\textbf{#3}}}}

\newsavebox{\astrutbox}
\sbox{\astrutbox}{\rule[-5pt]{0pt}{20pt}}

\begin{document}

\title{Sources and fluxes of scale-energy in the overlap layer of wall turbulence}
\author{A. Cimarelli, E. De Angelis, P. Schlatter, G. Brethouwer, A. Talamelli and C.M. Casciola}%

\date{}

\maketitle

\begin{abstract}
Direct Numerical Simulations of turbulent channel flows at friction Reynolds number 550, 1000, 1500, are
used to analyse the turbulent production, transfer and dissipation mechanisms in the compound space of scales
and wall-distances by means of the Kolmogorov equation generalized to inhomogeneous anisotropic flows. Two distinct 
peaks of scale-energy source are identified. The first stronger one belongs to the near-wall cycle. 
Its location in the space of scales and physical space is found to scale in viscous units while its intensity grows 
slowly with $Re$, indicating a near-wall modulation. The second source peak is found further away from the wall 
in the putative overlap layer and it is separated from the near-wall source by a layer of significant scale-energy sink. The 
dynamics of the second outer source appears to be strongly dependent on the Reynolds number. The detailed scale-by-scale 
analysis of this source highlights well-defined features that are used to make the properties of the 
outer turbulent source independent of Reynolds number and wall-distance by rescaling the problem. Overall, the 
present results suggest a strong connection of the observed outer scale-energy source with the 
presence of an outer region of turbulence production  whose  mechanisms are well separated from the near-wall region and 
whose statistical features agree with the hypothesis of an overlap layer dominated by attached eddies. Inner-outer 
interactions between the near-wall and outer source region in terms of scale-energy fluxes are also analysed. It is conjectured 
that the near-wall modulation of the statistics at increasing Reynolds number can be related to a confinement of 
the near-wall turbulence production due to the presence of increasingly large production scales in the outer scale-energy source region.

\end{abstract}

\section{Introduction}

One of the most peculiar aspects of turbulence in wall bounded flows is the ability of the turbulent
fluctuations to regenerate themselves through self-sustained processes. In wall flows, 
the production of turbulent fluctuations is embedded in the system instead of being provided by an external
agent. The dynamics of these self-sustaining mechanisms has been extensively investigated over the past
thirty years, since these processes are responsible for the energy drain from the mean flow to the
fluctuating field and for the turbulent drag. 

It has long been understood that the near-wall layer, being the site of the highest rate of turbulent 
energy production and of the maximum turbulent intensities, is crucial to the dynamics of attached shear flows.
The possibility to identify robust kinematic features in the proximity of a wall fed the hope of the scientific 
community to obtain a complete and consistent dynamical description of the underlying physics of these processes. 
The turbulent fluctuations near a wall have been found to organize in well defined coherent motions consisting of 
quasi-streamwise vortices and high/low velocity regions alternating in the spanwise direction. The former are 
longitudinal vortices with typical streamwise 
and spanwise length scales $\lambda_x^+ \approx 200$ and $\lambda_z^+ \approx 50$, respectively (hereafter a 
superscript + will denote the so-called inner units, see e.g. \citet{Townsend}), slightly tilted away from the 
wall. The latter are long and wide alternating arrays of streamwise streaks of local velocity excess/defect, with 
length scales $\lambda_x^+ \approx 1000$ and $\lambda_z^+ \approx 100$, superimposed on the mean flow. These 
features have been recognized in several numerical and experimental works, see e.g. \citet{Kim_Kline_Reynolds, Smith} 
and \citet{Robinson}. From these observations, several scientists tried
to derive a conceptual model of these processes. Following the work of \citet{Jimenez_autonomous}, the continuous 
creation and destruction of these turbulent structures form a self-sustaining cycle maintaining near-wall turbulence 
without the need of any input from the core flow, i.e. it is an autonomous cycle. The streamwise vortices extract 
energy from the mean flow to create alternating streaks of longitudinal velocity. Presumably by inflectional 
instabilities, these streaks in turn give rise to the vortices closing the cycle, see also \citet{Waleffe, Hussain, Schoppa}.

From a more applied point of view, the near-wall cycle is crucial since it controls the magnitude of the wall
stress. But coherent structures exist also at larger scales in the so-called overlap layer
\citep{Marusic, Hutchins, Jimenez}, and have been recently suggested to form an outer self-sustaining mechanism of regeneration
of very large turbulent fluctuations, see e.g. \citet{Flores, Mizuno, Cossu}. The phenomenology resembles the 
self-regenerating cycle near the wall though its characteristic dimensions are larger, see e.g. \citet{Monty,delAlamo}. 
The coherent motions involved in this outer cycle should scale with external variables meaning that their dimensions and action 
should increase as the extent of the log-layer widens with Reynolds number. Hence, the understanding of these outer 
dynamics is crucial for the modeling of wall-turbulence in the asymptotic regime of very large Reynolds number.
Furthermore, its analysis could help to clarify the interactions between the outer and inner regions 
of wall flows needed in the formulation of near-wall models for LES, see e.g. \citet{Piomelli}, and to explain the controversial 
mixed inner/outer scaling of the near-wall quantities such as spectra and turbulent intensities, see e.g. 
\citet{Marusic, DeGraaff}.

Generally speaking, the problem of wall-turbulent flows has been classically studied by
dividing the flow domain into well characterized regions depending on wall-distance. In particular,
wall-bounded flows are divided in a near-wall, inner region, and an outer region populated
by large structures. These two distinct regions are present in all wall-bounded flows and interact in the overlap region. 
While in the outer flow the velocity profile depends on the particular
flow configuration, in the inner and overlap regions it exhibits a high degree of universality starting linearly from the wall and then approaching a logarithmic behaviour. These behaviors opportunely scaled with viscous
units should collapse for different flows and different Reynolds numbers, see \citet{Nagib} for a detailed description 
of the controversies on this topic. The same scaling should apply to
the turbulent intensity profiles and to all the statistical observables of the inner region.

However, the near wall quantities exhibit a Reynolds dependence as shown by the fact that the energy of the long 
turbulent fluctuations of the overlap layer grows when the Reynolds number increases. This large-scale motion is 
found to actively modulate the near-wall turbulence by production of near-wall scales at increasing 
Reynolds number \citep{Marusic}. The observed increase of the streamwise turbulent fluctuation peak, the possible 
appearance of a second peak in the overlap flow and the presence of a marked outer-scale peak in the energy spectrum
are thought to be a signature of these effects. An important consequence of the Reynolds number 
dependence of the large turbulent motion in the overlap layer is that most of the turbulence production should 
asymptotically come from this region due to the widening of the overlap layer with $Re$ \citep{SmithARF}. Even if no 
Reynolds dependence for the outer turbulent production intensity is expected, the outer turbulent self-sustained 
mechanisms are thought to dominate the high Reynolds number asymptotic state of wall turbulence.

Given the Reynolds dependence of these processes in the compound scale/physical-space,
an interesting approach to study the basic mechanisms of the outer cycle has been recently proposed in \citet{Cimarelli_JFM2}
by extending the statistical approach used in \citet{Marati} and applied by \citet{Neela} to moderately high Reynolds number data. 
The classical approach for addressing these issues in the channel flow is based on a Fourier decomposition along the homogeneous directions  while keeping a description in terms of the physical distance in the wall normal direction, see e.g. \citet{Jimenez_spectra}.  However this more traditional approach does not allow for a net distinction between position in the wall normal direction and wall normal scale at which energy generation and energy flux take place. The tool used here to describe the energy content associated with a given scale of motion in a given position in space is based, instead,  on the generalization of the Kolmogorov equation for the second order structure function, originally introduced for homogenous and isotropic turbulence, and successively extended to inhomogenous, anisotropic flows by \citet{Hill}. The generalized Kolomogorov equation  keeps the two concepts of wall normal position and wall normal scale clearly distinct, thereby allowing to distinguish between the two associated components of the energy flux.
In \citet{Cimarelli_JFM2}, this multi-dimensional description of turbulence has been used and proven fundamental for the understanding
of the wall-turbulent physics and for its modeling as shown in \citet{Cimarelli_JFM}. In the present paper, we 
extend this work by analysing how the turbulent energy associated to a certain scale (scale-energy) is generated, transferred and dissipated among different scales 
and wall-distances varying the Reynolds number with particular attention to the outer self-regeneration mechanisms. 

\section{DNS database and single point statistics}

\begin{table}
\begin{center}
\begin{tabular}{*{8}{c}}

Case & $Re_{\tau}$  &  $L_x$  &  $L_y$  &  $L_z$ & $N_x \times N_y \times N_z $ & $\Delta x^+$ & $\Delta z^+$ \\ \\

DNS550 & $550$  &  $8\pi h$  &  $2h$  &  $4\pi h$  & $1024 \times 257 \times 1024$ & $13.5$ & $6.7$ \\
DNS1000 & $1000$  &  $8\pi h$  &  $2h$  &  $3\pi h$  & $2560 \times 385 \times 1920$ & $9.8$ & $4.9$ \\
DNS1500 & $1500$  &  $12\pi h$  &  $2h$  &  $10.5 h$  & $6144 \times 577 \times 3456$ & $9.2$ & $4.5$ \\

\end{tabular}
\end{center}
\caption{Parameters of the simulations. $Re_\tau$ is the friction Reynolds number, $L_x$, $L_y$ and $L_z$ are the lengths 
of the computational domain in the streamwise ($x$), wall-normal ($y$) and spanwise ($z$) direction,  $N_x$, $N_y$ and $N_z$
are the number of points in physical space and  $\Delta_x^+$, $\Delta_y^+$ and $\Delta_z^+$ the corresponding grid spacing in 
viscous units.}
\label{simulations}
\end{table}

In the present study, we analyze data of three direct numerical simulations (DNS) of fully developed
turbulent channel flow at $Re_\tau = u_\tau h / \nu = 550$, $1000$ and $1500$
respectively. Here, $u_\tau$ is the friction velocity, $h$ the channel half gap width
and $\nu$ the viscosity. The simulations were carried out with a pseudo-spectral code using 
Fourier expansions and dealiasing in the homogeneous directions, and Chebyshev polynomials 
in the wall-normal direction. Full details of the algorithm can be found in \citet{Chevalier}.
The domain size and resolution of the three DNSs are given in table \ref{simulations}.
Data of the DNSs at $Re_\tau = 550$ and 1000 have already been used for
studies of wall-turbulence in \citet{Cimarelli_JFM2} and \citet{Lenaers}, respectively. Let us mention 
that the lower resolution adopted for the simulation of the lower Reynolds number case at $Re_\tau = 550$, has 
been tested and found to not affect the statistical quantities we are analysing in the present work.

\begin{figure}
\begin{center}
\includegraphics[width=0.45\linewidth]{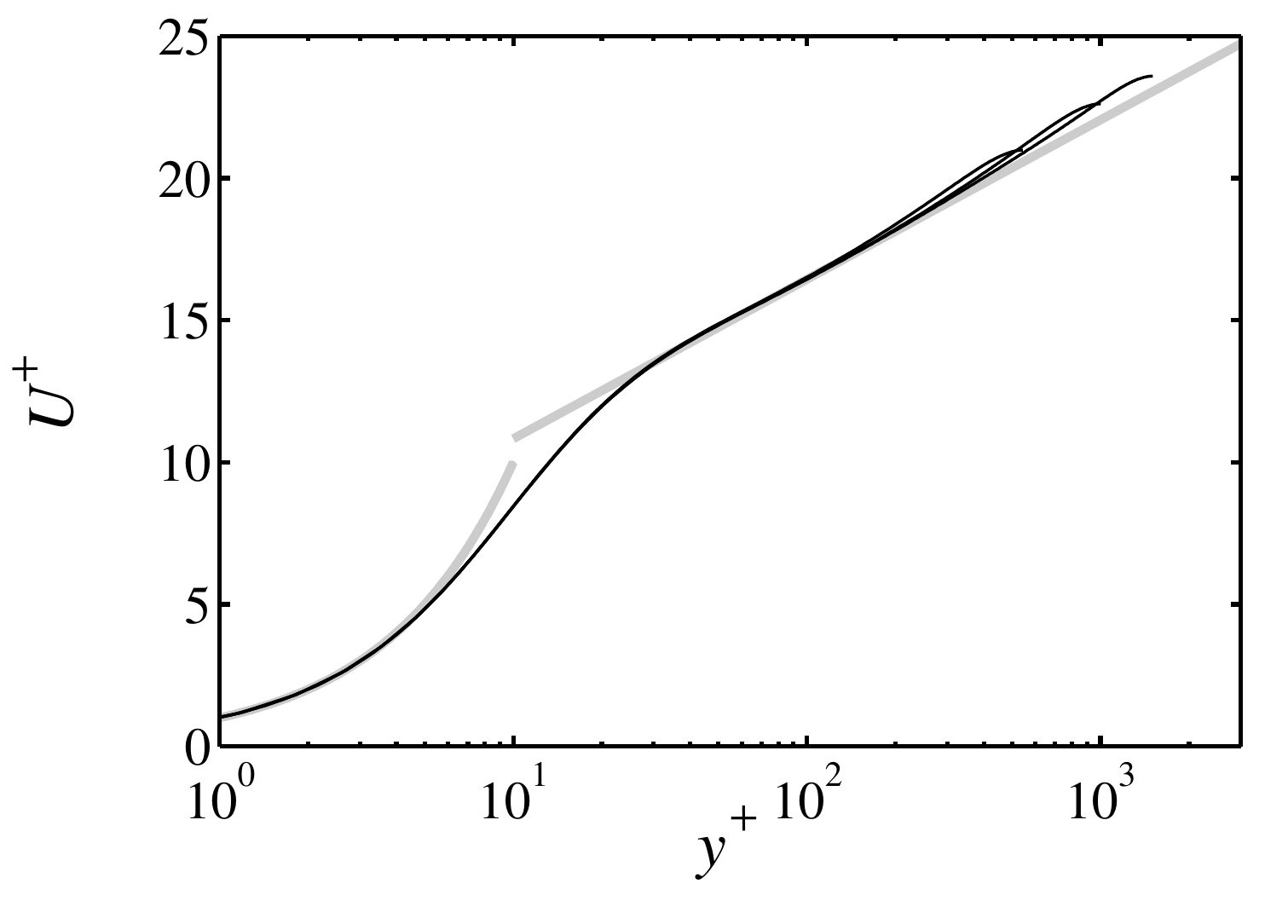}%
\mylab{-49mm}{35mm}{(\aaa)}%
\includegraphics[width=0.45\linewidth]{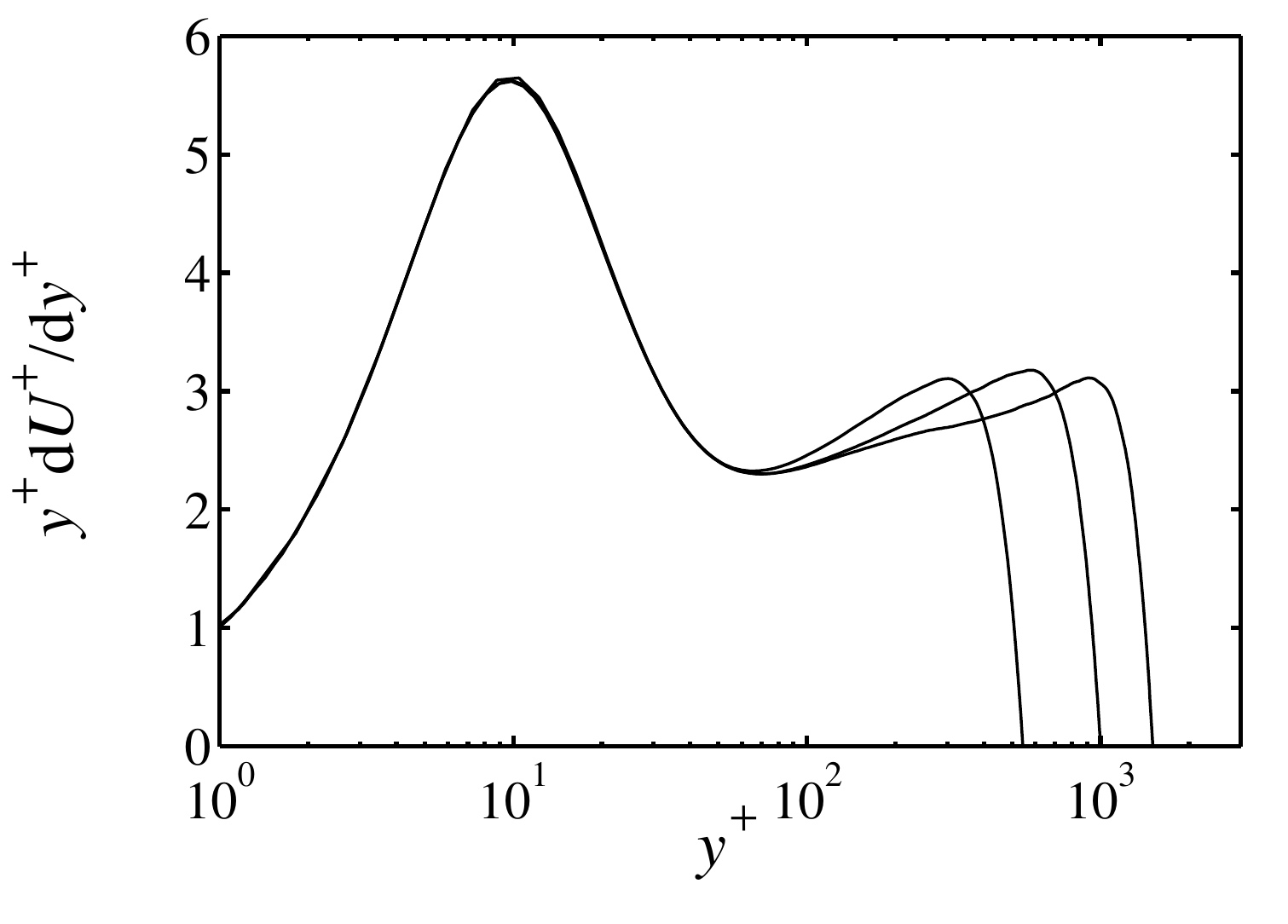}%
\mylab{-49mm}{35mm}{(\bbb)}%
\caption{Mean velocity profiles $U^+(y^+)$ and (premultiplied) mean velocity gradient
$y^+ \mathrm{d}U^+/\mathrm{d}y^+$ for the three channel DNS considered in
this study, with $Re_\tau=550$, 1000, 1500. $U^+(y^+) = y^+$ and $U^+(y^+) = 1/\kappa \log(y^+) +  B$ with 
$\kappa=0.41$ and $B=5.2$ are indicated by thick gray lines.}
\label{umean}
\end{center}
\end{figure}

Profiles of the streamwise mean velocity and the log-law indicator function of the
three DNSs are shown in figure \ref{umean}. The near-wall region has obviously a very high
degree of similarity for the three Reynolds numbers, while outside the buffer layer, in the overlap
layer, differences become visible due to Reynolds number effects. A tentative plateau
in the indicator function profile starts to appear with increasing $Re_\tau$ but
is not yet clearly present, meaning that a true logarithmic layer is
absent even at the highest $Re_\tau$.

\begin{figure}
\begin{center}
\includegraphics[width=0.49\linewidth]{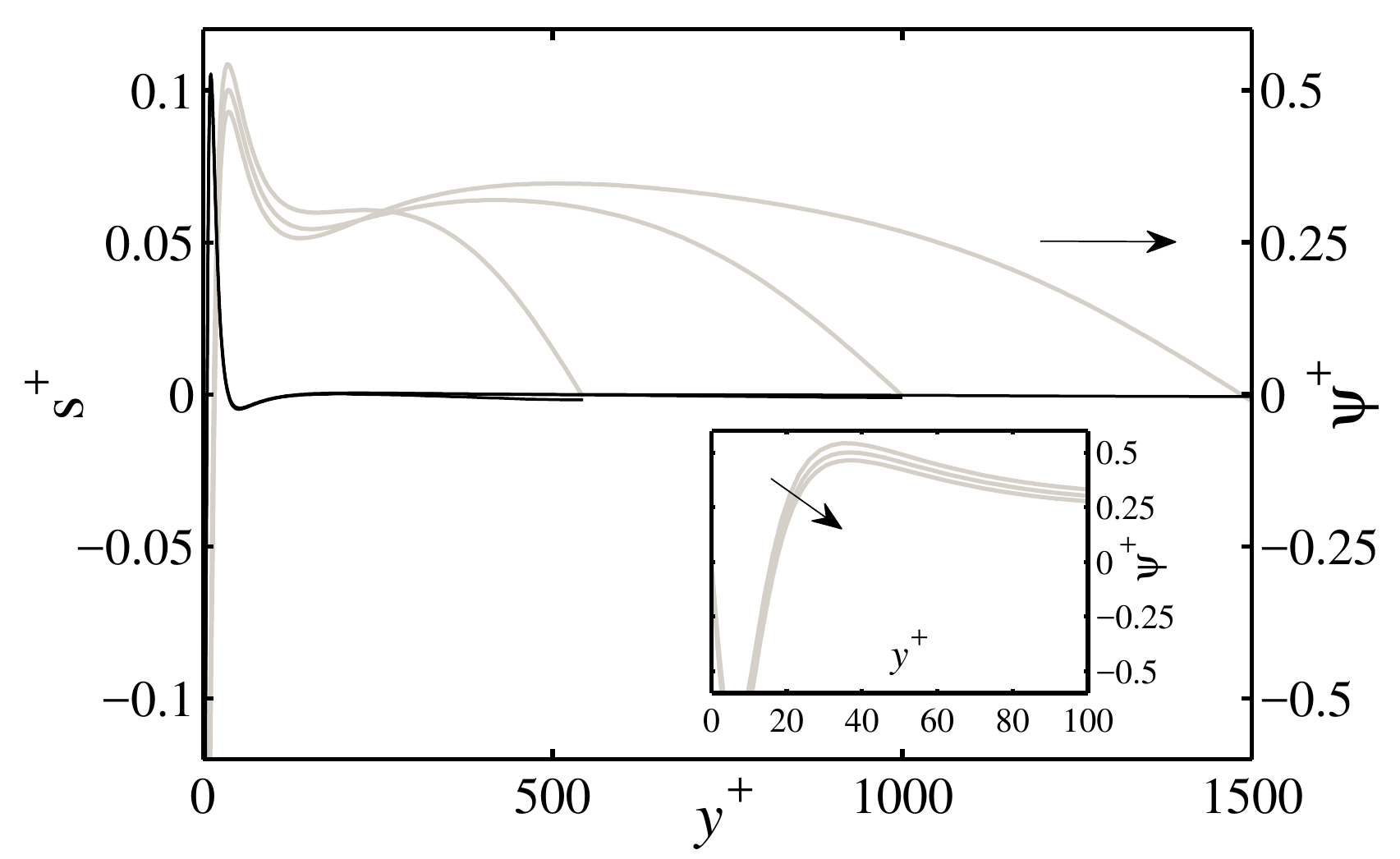}%
\mylab{-16mm}{35mm}{(\aaa)}%
\includegraphics[width=0.46\linewidth]{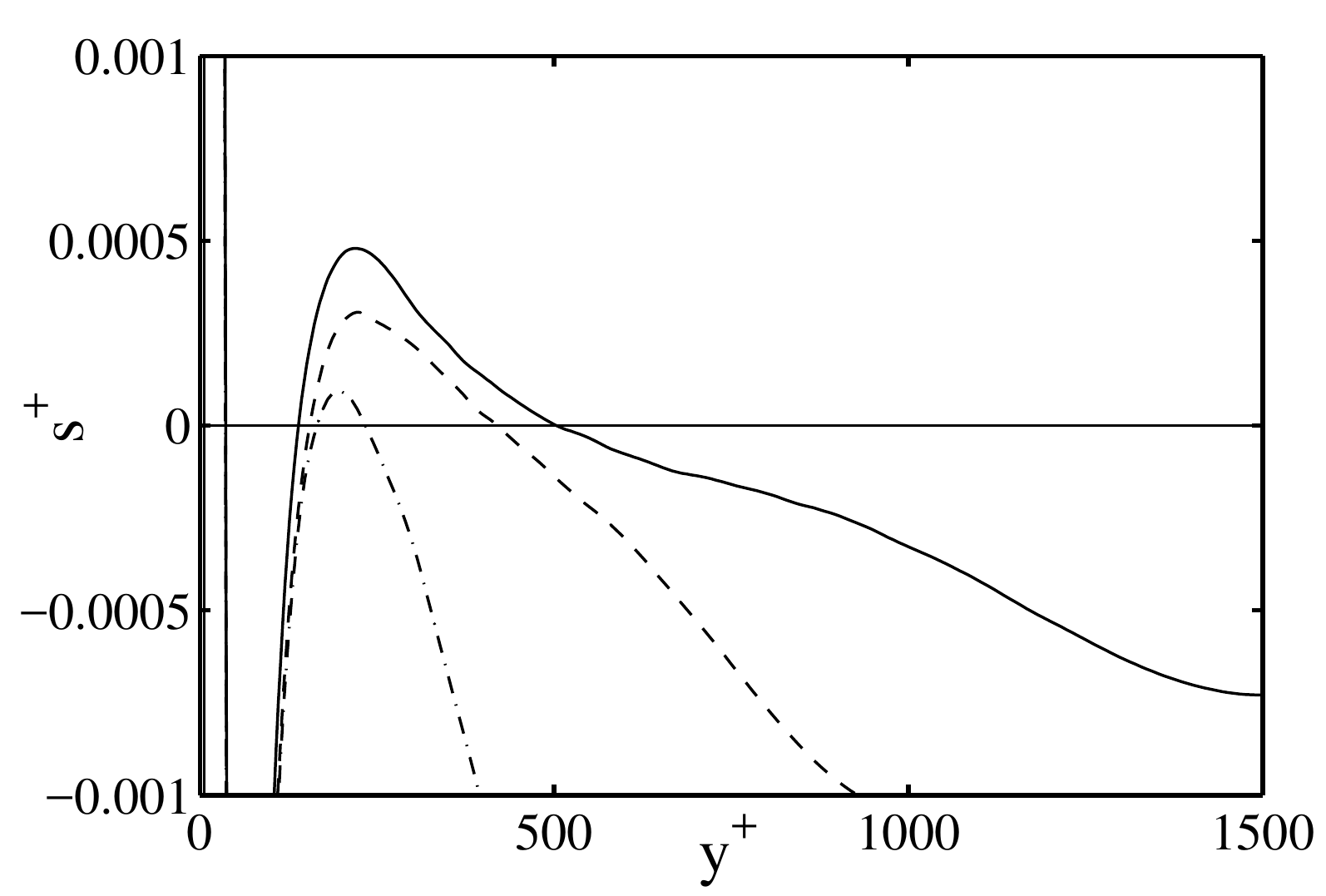}%
\mylab{-16mm}{35mm}{(\bbb)}%
\caption{(a) Source term of the kinetic energy budget $s^+$ (black lines) and corresponding spatial flux
$\psi^+$ (grey lines) for the three Reynolds numbers $Re_\tau=550$, 1000, 1500. The inset shows a zoom of the near-wall peak of the spatial flux $\psi^+$. The two arrows indicate increasing Reynolds numbers. (b) Magnification of the region around the second positive peak of $s$. Increasing $Re_\tau$ from dash-dotted to dashed and solid lines.}
\label{tke}
\end{center}
\end{figure}

The single-point turbulent kinetic energy balance compacts in a simple way the overall 
multidimensional behavior of turbulence describing the energetics only in physical-space. 
For the symmetries of the channel, this equation reads,
\begin{equation}
\frac{d \psi}{dy} = s(y) \; ,
\label{TKE_channel}
\end{equation}
and describes how turbulent energy is redistributed among different wall-distances $y$ through
the spatial flux 
$\psi = ( \langle{u_i^2 v}\rangle +  \langle{p v}\rangle / \rho -\nu d  \langle u_i^2 \rangle / dy)$
from the production to the dissipation regions of the flow defined by positive and negative values of 
the  source term 
$s(y)= -  \langle{u v}\rangle (dU / dy) - \langle \epsilon \rangle$. Hereafter, $\langle \cdot \rangle$ will be used
to denote ensamble average.
As shown by the black lines in figure \ref{tke}(a), in wall-turbulence the energy source is near the wall 
in the so-called buffer layer. In this layer turbulence production exceeds the local dissipation. Conversely, the
wall and bulk flow behave as sink regions dissipating turbulent energy emerging from the buffer layer through the spatial
energy flux, see gray lines in figure \ref{tke}(a). Indeed, the spatial flux is zero at the peak of energy 
source and becomes positive (towards the core flow) further away from the wall and negative (towards the wall) closest to 
the wall. Actually, in between the buffer layer and the core of the flow, a third region can be defined, the so-called 
overlap layer, which is the main subject of the present work. Although this region is expected
to be an equilibrium layer where production and dissipation locally balance, production is actually larger than dissipation 
leading to an outer energy source, see figure \ref{tke}(b). The understanding of this region of the flow is very 
important especially when dealing with the large Reynolds number state of wall-turbulence. Even if its intensity
is very small compared to the one near the wall, this outer energy source shows an apparent $Re$-dependence, 
see again figure \ref{tke}(b). In particular, it appears that by increasing the Reynolds number, the role of 
the outer energy source becomes more important. By defining the overall inner and outer energy source 
as the integral of the source $s(y)$ restricted to the two (inner and outer) regions where $s(y) > 0$, respectively,
$$
\Xi_{inn} = \int_y s(y) dy  \qquad y \in \{\mbox{inner region of energy source} \; s(y) >0\} 
$$
$$
\Xi_{out} = \int_y s(y) dy  \qquad y \in \{\mbox{outer region of energy source} \; s(y) >0\} \; ,
$$
we can roughly estimate the relative importance of the two regions as function of Reynolds number by means of the ratio $\Xi_{out}/\Xi_{inn}$. The present data show a significant increase of this ratio
from $0.0031$ at $Re_\tau =550$ to $0.0347$ and $0.0617$ at $Re_\tau = 1000$ and $1500$, respectively. Extrapolating this trend, one can expect the outer source to become dominant above $Re_\tau = 15000 \div 20000$.
The increased intensity of the outer source has also consequences for the topology of the 
energy transfer. Indeed, as shown in figure \ref{tke}(a), the spatial flux in the overlap layer increases 
with $Re$ and forms an outer peak given by the increasing energy injection due to the outer energy source.

\section{Generalized Kolmogorov equation}

The generalized Kolmogorov equation proposed by \citet{Hill} is the balance equation for the second order structure 
function, $\langle \delta u^2 \rangle$, where $\delta u^2 = \delta u_i \delta u_i$ and the fluctuating velocity 
increment is $\delta u_i = u_i(X_s+r_s/2)-u_i(X_s-r_s/2)$. According to its definition, 
$\langle \delta u^2 \rangle (r_i, X_i)$ depends both on the separation vector defined as $r_i = x_i^\prime - x_i$ and on the
location specified by the mid-point $X_i = (x_i^\prime + x_i )/2$.
Hereafter, index repetition implies summation. The second order structure function can be interpreted as
the amount of energy of a given scale $r_s$ at a certain position in the flow $X_s$ and for that reason, hereafter,
we will refer to the concept of scale-energy. Let us note that, although 
$\langle \delta u^2 \rangle$ has the dimensions of kinetic energy and it is strictly related to the energy spectrum, 
the second order structure function is not an intensive quantity. However, it represents the natural tool
for the multi-scale analysis of turbulent flows that lack a classical spectral decomposition due to violation
of spatial homogeneity. The second order structure function is defined in a four-dimensional space $(r_x,r_y,r_z,Y_c)$
allowing to distinguish fluxes between different wall-distances $Y_c$ and fluxes between different wall-normal scales $r_y$. This 
distinction would be missed by using the balance equation for spectral energy. The generalized Kolmogorov equation derives directly 
from the Navier-Stokes equations. For the symmetries of channel flow and considering increments $r_s$ only 
in the directions parallel to the walls, $r_y=0$, \citep{Marati}, the equation reads,
\begin{eqnarray}
\frac{\partial \langle \delta u^2 \delta u_i \rangle}{\partial{r_i}} +
2 \langle \delta u \delta v \rangle \left ( \frac{dU}{dy} \right )^* +
\frac{\partial \langle v^* \delta u^2 \rangle}{\partial Y_c} 
& = &  \nonumber \\
- 4\langle \epsilon^* \rangle +
2\nu \frac{\partial^2 \langle \delta u^2 \rangle}{\partial{r_i} \partial{r_i}} -
\frac{2}{\rho} \frac{\partial \langle \delta p \delta v \rangle}{\partial Y_c} 
& + &
\frac{\nu}{2} \frac{\partial^2 \langle \delta u^2 \rangle}{\partial {Y_c}^2}\ .
\label{Hill_eq}
\end{eqnarray}
where $U(y)$ is the mean velocity profile, $Y_c = X_2$ is the wall-normal coordinate of the mid-point,
(*) denotes the arithmetic average at the points  $X_s \pm r_s/2$ and 
$\epsilon = \nu (\partial u_i / \partial x_j) (\partial u_i / \partial x_j)$ is the pseudo-dissipation.
Equation~(\ref{Hill_eq}) involves a 
four-dimensional vector field,  $\mathbf{\Phi}=(\Phi_{r_x},\Phi_{r_y},\Phi_{r_z},\Phi_c)$, and can be restated as
\begin{equation}
\nabla_4 \cdot \mathbf{\Phi} (\mathbf{r},Y_c )  = 
\xi(\mathbf{r},Y_c)\, ,
\label{conservation_form}
\end{equation}
where $\nabla_4$ is the four-dimensional gradient and
$\xi = -2 \langle \delta u \delta v \rangle \left ( dU / dy \right )^* - 4\langle \epsilon^* \rangle$ is
the scale-energy source/sink given by the balance between production and dissipation. This equation allows to 
identify the two transport processes occurring simultaneously in wall-flows: the scale-energy transfer in 
the three-dimensional space of scales,
$\mathbf{\Phi}_r = (\Phi_{r_x},\Phi_{r_y},\Phi_{r_z}) = 
\langle \delta u^2 \delta \mathbf{u}\rangle - 2 \nu \nabla_r \langle \delta u^2 \rangle$ and the spatial energy flux among 
different wall-distances,
 $\Phi_c = \langle v^* \delta u^2 \rangle + 2 \langle \delta p \delta v \rangle / \rho - 
\nu/2 \partial \langle \delta u^2 \rangle / \partial Y_c$.  In the inertial sub-range of homogeneous isotropic turbulence, 
eq.(\ref{conservation_form}) reduces to 
\begin{equation}
\nabla_3 \cdot \mathbf{\Phi}_r (\mathbf{r})  = - 4\langle \epsilon \rangle \, ,
\label{Hom_conservation_form}
\end{equation}
where the contributions due to production, viscosity and spatial inhomogeneity are either negligible or zero.
In this case, energy transport  occurs only in the space of scales, is radial and from large to small scales.  
The scale-energy source, $\xi_{hom} =  \langle \delta u \delta f \rangle - 4\langle \epsilon \rangle$ where $f$ is 
the external forcing, is a function only of the separation vector modulus $|\mathbf{r}|$ and it is always 
negative, $\xi_{hom} = g(|\mathbf{r}|) \leq 0$. In inhomogeneous flows, the turbulent production can locally exceed 
dissipation leading to regions of scale-energy source in the augmented $(\mathbf{r},Y_c)$-space where $\xi(\mathbf{r},Y_c)>0$. 
This is a distinguishing feature of actual inhomogeneous flows that has been shown responsible in \citet{Cimarelli_JFM2} for 
a complex redistribution of scale-energy where the controversial reverse energy cascade plays a central role. Hence, in what 
follows, the study of the Reynolds number effects on the energetics of the flow will mainly focus on the behavior of the 
source term $\xi$.

\begin{figure}
\begin{center}
\includegraphics[width=0.3\linewidth]{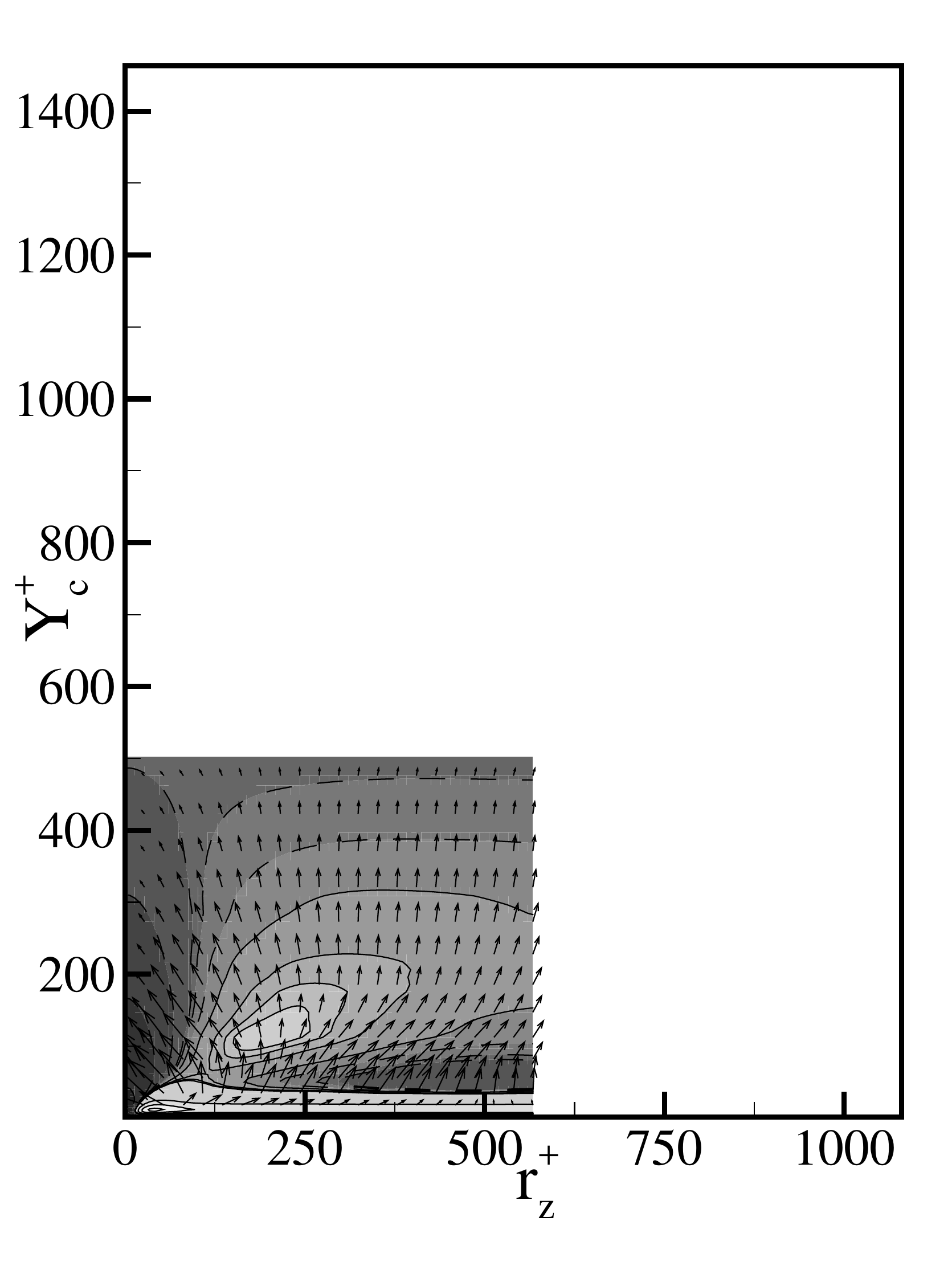}%
\mylab{-33mm}{45mm}{(\aaa)}%
\includegraphics[width=0.3\linewidth]{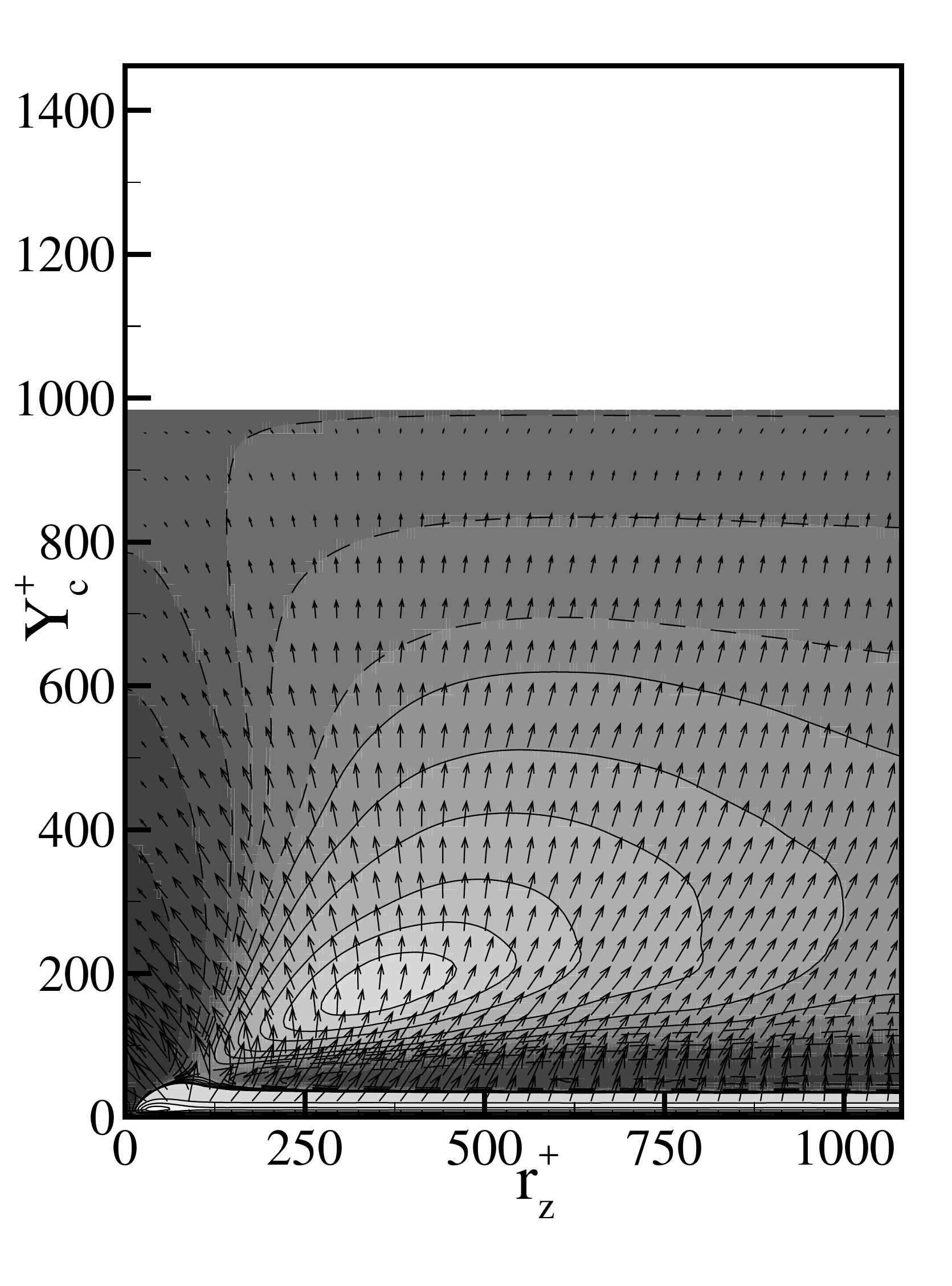}%
\mylab{-33mm}{45mm}{(\bbb)}%
\includegraphics[width=0.3\linewidth]{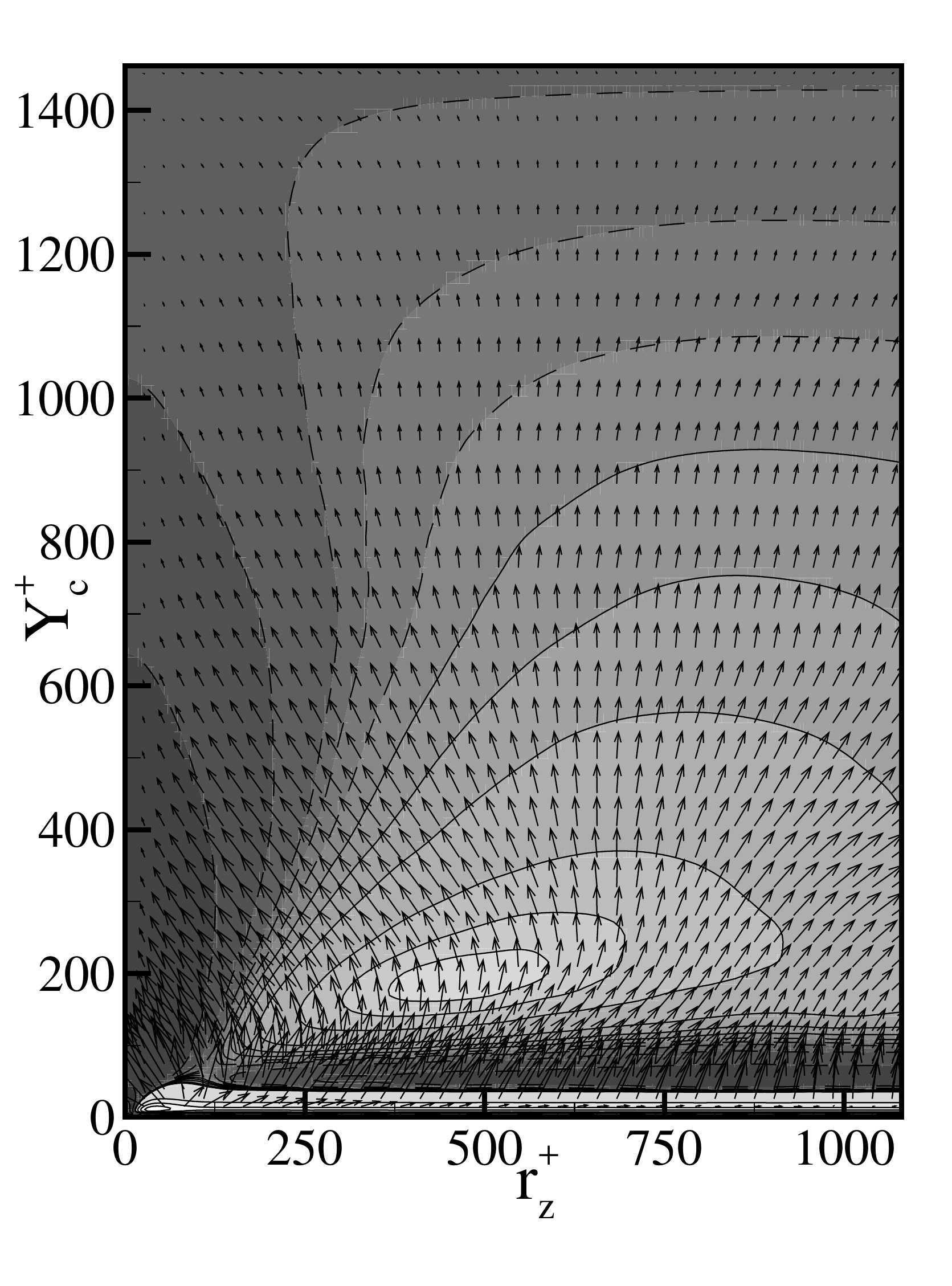}%
\mylab{-33mm}{45mm}{(\ccc)}%
\caption{Direct numerical simulation of turbulent channel flows at $Re_\tau=550$, $Re_\tau=1000$ and $Re_\tau=1500$ from left to right.
Energy source -- $\xi^+(r_x,r_z,Y_c)$  -- isolines and field of fluxes -- $(\Phi_{r_z}^+, \Phi_c^+)$ -- vectors in the $r_x=0$ plane.
Energy source intensity increases from black to white colors. Solid lines denote postive values while dashed lines denote 
negative values.}
\label{rx0_behaviors}
\end{center}
\end{figure}

\section{The structure of the source term}
 
The topological structure of the source term obtained from the three DNS data sets shows that all the basic
characteristics observed at the lowest $Re_\tau$ are also maintained at higher Reynolds numbers. 
By analyzing the data in the reduced space $(r_x,r_z,Y_c)$ for $r_y=0$, the source $\xi$ is found to reach 
its maximum at $r_x=0$ in a range of small spanwise scales well within the buffer layer, see figure \ref{rx0_behaviors}. 
This region of the reduced space is a singularity point for the fluxes. As more clearly seen in figure \ref{NW_scaling},
the scale-energy flux vector field takes origin from this peak of scale-energy source which can be considered as 
the engine of wall-turbulence and will be hereafter called the driving scale range (DSR). The small-scale location 
of the scale-energy source actually violates the classical paradigm of homogeneous isotropic turbulence. In wall flows, 
the turbulent energy is generated amid the spectrum of turbulent fluctuations, not at the largest scales, and this fact leads 
to a complex redistribution of energy \citep{Cimarelli_JFM2}, with strong consequences for turbulence modeling 
\citep{Cimarelli_PoF, Cimarelli_PoF2}.

Another interesting feature emerging from the analysis of the Kolmogorov equation is the existence of a rescaled
replica of the DSR, associated with a second peak in the scale-energy source, called hereafter outer driving scale 
range (ODSR). This outer peak of scale-energy source has been observed in \citet{Cimarelli_JFM2} and is present 
also at higher Reynolds numbers, see the isocontours of figure \ref{rx0_behaviors}.

\subsection{The driving scale range DSR}

\begin{figure}
\begin{center}
\includegraphics[width=0.7\linewidth]{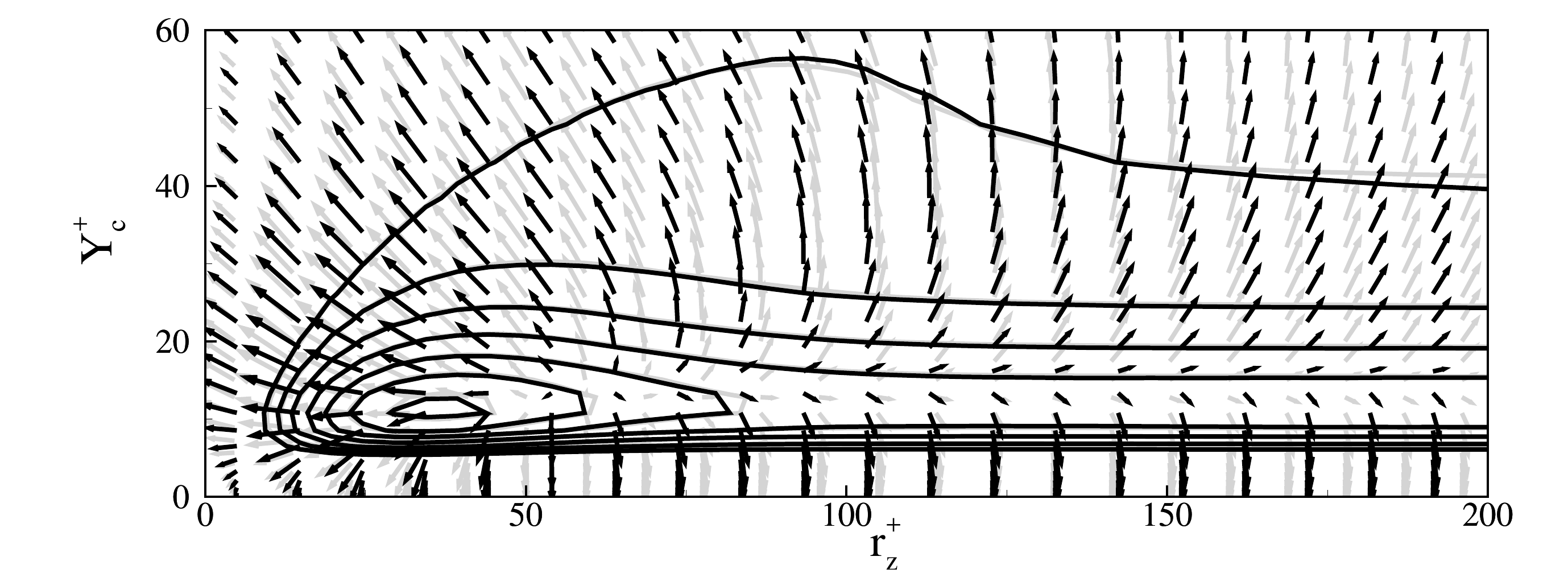}
\caption{Scaling of the near-wall scale-energy source ($\xi^+$) and of the field of fluxes -- $(\Phi_{r_z}^+, \Phi_c^+)$ 
in the $r_x=0$ plane. Grey isolines and vectors denote the $Re_\tau =1000$ case while black ones denote the $Re_\tau =1500$ case.}
\label{NW_scaling}
\end{center}
\end{figure}

\begin{figure}
\begin{center}
\includegraphics[width=0.59\linewidth]{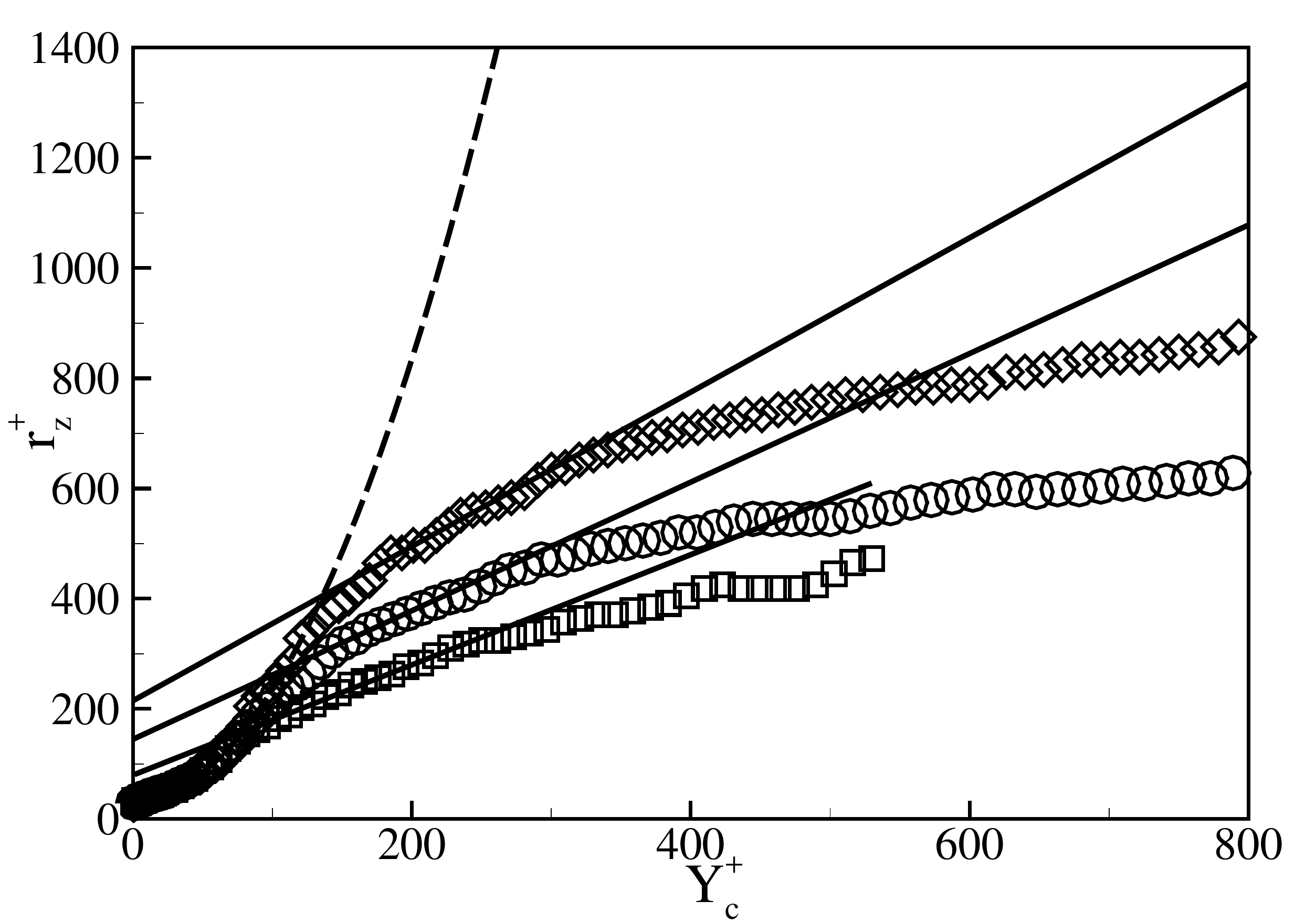}%
\mylab{-68mm}{45mm}{(\aaa)}%
\includegraphics[width=0.4\linewidth]{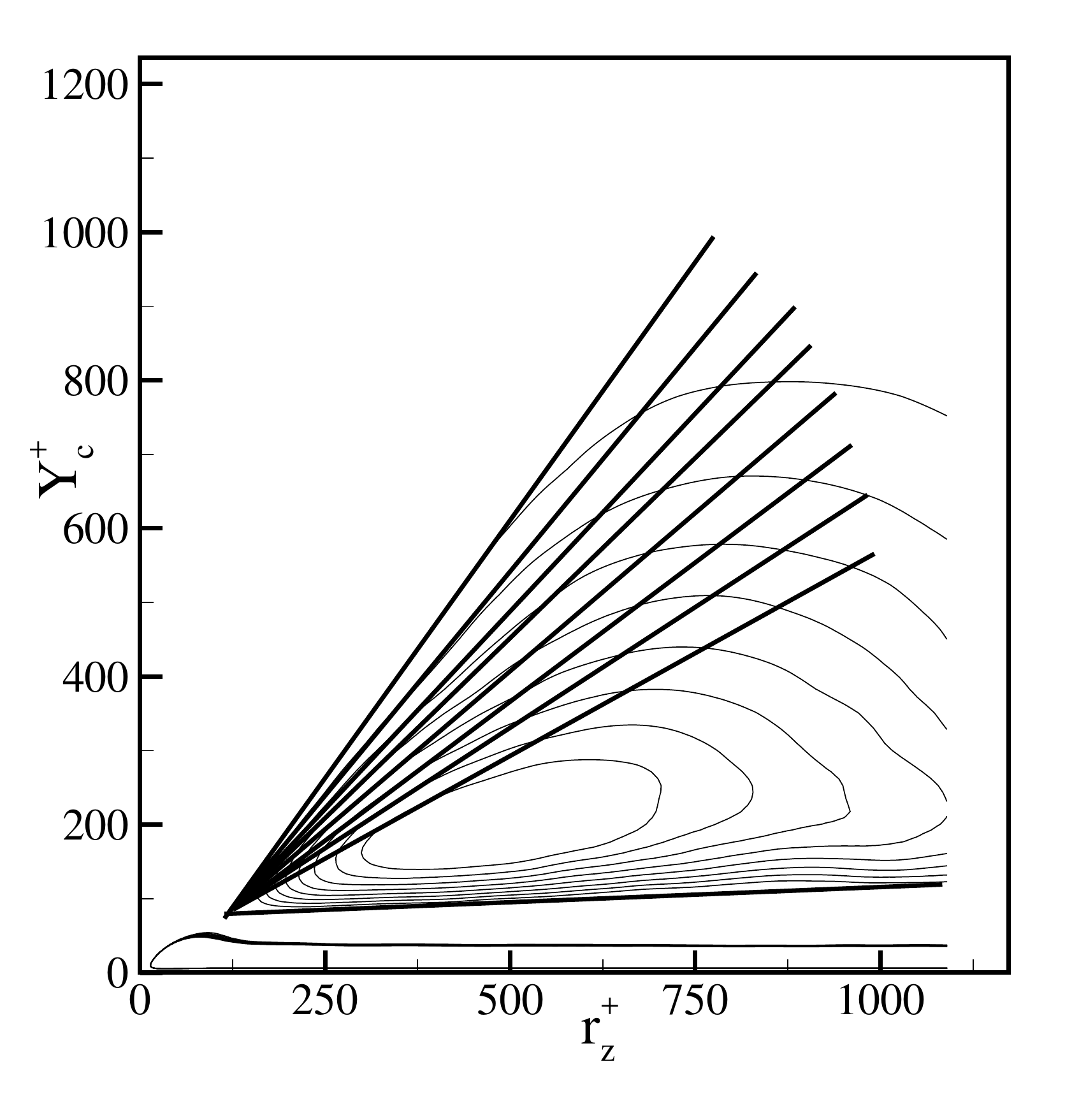}%
\mylab{-43mm}{45mm}{(\bbb)}%
\caption{(a) Spanwise scale of maximum scale-energy source, $(\ell_z^{\xi_{max}})^+$, as a function of the wall distance. 
Diamonds denote the $Re_\tau =1500$ case, circles the $Re_\tau =1000$ case and squares the $Re_\tau =550$ case.
The dashed line denotes the quadratic increase near the wall, $(\ell_z^{\xi_{max}})^+ = 35 + 0.02 Y_c^{+2}$ while 
the solid lines the linear increase in the overlap layer, $(\ell_z^{\xi_{max}})/h = 0.14 + (Y_c/h)$, $(\ell_z^{\xi_{max}})/h
= 0.14 + 1.16 (Y_c/h)$ and $(\ell_z^{\xi_{max}})/h = 0.14 + 1.4 (Y_c/h)$, expressed in viscous units, i.e. 
$(\ell_z^{\xi_{max}})^+ = 80 + Y_c^+$, $(\ell_z^{\xi_{max}})^+ = 145 + 1.16 Y_c^+$ and $(\ell_z^{\xi_{max}})^+ = 
215 +1.4 Y_c^+$. (b) Energy source -- $\xi^+$  -- isolines in the $r_x=0$ plane for the $Re_\tau=1500$
case. The straight lines are $r_z^+ = \gamma (\xi_0^+) (Y_c^+ - \tilde{Y}_c^+) + \tilde{r}_z^+$ where $\xi_0^+$ defines the 
iso-level of $\xi$, $\gamma (\xi_0^+) \in [0.04 , 1.4]$, $\tilde{r}_z^+ = 120$ and $\tilde{Y}_c^+ = 80$.}
\label{scaling}
\end{center}
\end{figure}

In agreement with the picture of a universal near-wall region, the geometrical properties of the DSR are 
unaffected by the Reynolds number, see figure \ref{NW_scaling}. For the three cases considered, 
the source peak within the DSR is located at $(r_x^+,r_z^+,Y_c^+)=(0,40,12)$. This Reynolds-number
invariance and the clear matching of scales and positions suggests a strong connection with the near-wall cycle. 
Note that for a given wall-distance the source maximum 
$\xi$ occurs at $r_x = 0$. Its location in the $(r_z,Y_c)$-plane, reported in figure \ref{scaling}(a), 
defines the typical spanwise scale of the scale-energy source, $(\ell_z^{\xi_{max}})^+$. Near the wall, the spanwise location 
of the maxima increases quadratically with the wall distance, $(\ell_z^{\xi_{max}})^+ \approx 35 + 0.02 Y_c^{+2}$,  
for all the Reynolds numbers considered. Clearly, within the DSR for small distances from the wall, $Y_c^+ < 30$, this 
trend results in an almost constant spanwise length scale: the typical spanwise scale of the most active structures of 
the wall is independent of the wall distance and Reynolds number. Although the topology of the DSR is basically $Re$-invariant, 
its intensity is not. The scale-energy source is found 
to slightly increase with $Re_\tau$. In particular we measure $\xi_{max}^+=0.717, 0.732, 0.741$ moving from the 
lower to the higher Reynolds number. This trend is consistent with the commonly observed mixed inner/outer 
scalings of the near-wall quantities. In fact, it is thought that the outer dynamics actively modulates the near-wall 
turbulence by producing small scale fluctuations increasing Reynolds number \citep{Marusic2010}.

As shown by the vector field in figure \ref{NW_scaling}, a direct consequence of the $Re$-invariance of the topology of the 
scale-energy source term, $\xi$, is that also the scale-energy flux vector field, $(\Phi_{r_z}, \Phi_c)$, 
remains identical for increasing Reynolds number, once rescaled in viscous units. In this scenario, the Reynolds-number 
effects should come only from the ODSR in the overlap region. Indeed, even if the observed second peak of scale-energy source $\xi$ is 
very small compared to the one in the DSR, its relevance increases with $Re$ as will be discussed in the next sections.

\subsection{The outer driving scale-range ODSR}
\label{OSR}

The ODSR belongs to the overlap layer and appears to be the result of a second outer turbulent production 
mechanism well separated from the near-wall dynamics. The ODSR is separated from the DSR by a
scale-energy sink region. Interestingly, this separation is found to 
be more pronounced by increasing $Re$. The solid contour lines shown in figure \ref{rx0_behaviors}, highlight that 
certain positive values for the source term $\xi$ are shared by both the DSR and ODSR for the $Re_\tau=550$. On the contrary,
for the larger Reynolds numbers, the DSR and ODSR are more and more separated by negative values for the source term $\xi$ as
shown by the number of dashed contours lines in between the inner and outer source shown in figure \ref{rx0_behaviors}.

Although the peak intensity is smaller than the DSR one, the extent of the ODSR increases with $Re$
suggesting how this object can play an important role at large Reynolds numbers. In contrast with the DSR where 
production is concentrated at small scales which are independent 
of Reynolds number, the ODSR involves larger scales and its extent in inner units increases with $Re_{\tau}$.
For the three Reynolds numbers considered, the peak of the ODSR occurs in a well-defined 
scale-region expressed in outer units corresponding to $r_z / h \approx 0.34$. On the other hand, the physical location of the 
peak in the ODSR is $Y_c / h=(0.2; 0.18; 0.12)$ in outer units while $Y_c^+=(112; 186; 192)$ in viscous units going from  
low to the high Reynolds number. Contrary to the space of scales, $r_z$, which is found to be 
$Re$-invariant once expressed in outer units, the wall distances, $Y_c$, do not scale with $Re$ neither in inner nor outer 
units. This behavior is probably related to the fact that the overlap layer extends from a lower 
limit given in viscous units to an upper limit in outer units, e.g. for $100 < Y_c^+ < 0.2 Re_\tau$ but 
the exact values are still a matter of scientific debate, see \citet{Marusic2010} and references therein. These arguments suggest 
a mixed scaling with wall-distance of the outer scale-energy source, see section \ref{scalings} for a more detailed 
discussion. When considering the ODSR intensity, it is worth noting that the peak of scale-energy source in the ODSR 
remains essentially unaltered with Reynolds number once scaled in inner units and corresponds to $\xi^+ \approx 0.0095$.

The presence of the ODSR violates the equilibrium assumption of the overlap layer from which a local balance of 
production and dissipation is expected. Within the ODSR the energy injection is larger than the rate of 
dissipation, $\xi > 0$. As stated by equation (\ref{conservation_form}), this fact results in a positive divergence of 
the energy transfer, $\nabla_4 \cdot \mathbf{\Phi} > 0$,  which under the assumption of a true equilibrium is otherwise 
expected to be zero. This observation
is consistent with the single-point energy excess already discussed in connection with figure \ref{tke} in 
section 2. Consequently, the overlap layer does not behave like a homogeneous shear flow traversed by a constant spatial 
flux of energy both at the single-point and two-point level. The ODSR continuously injects energy feeding the energy fluxes.
As shown in figure \ref{tke}, the spatial flux starts from the DSR in the buffer layer, it decreases by delivering energy 
in the sink layer wedged between the DSR and ODSR to increase again due to the energy injected by the ODSR leading to the 
second peak of the spatial flux. This second peak strongly depends on the Reynolds number, since the underlying physics
of the ODSR belongs to the overlap layer whose extent and, hence, its overall energy injection, increases with $Re$, 
as figure \ref{rx0_behaviors} clearly suggests. 

\begin{figure}
\begin{center}
\includegraphics[width=0.59\linewidth]{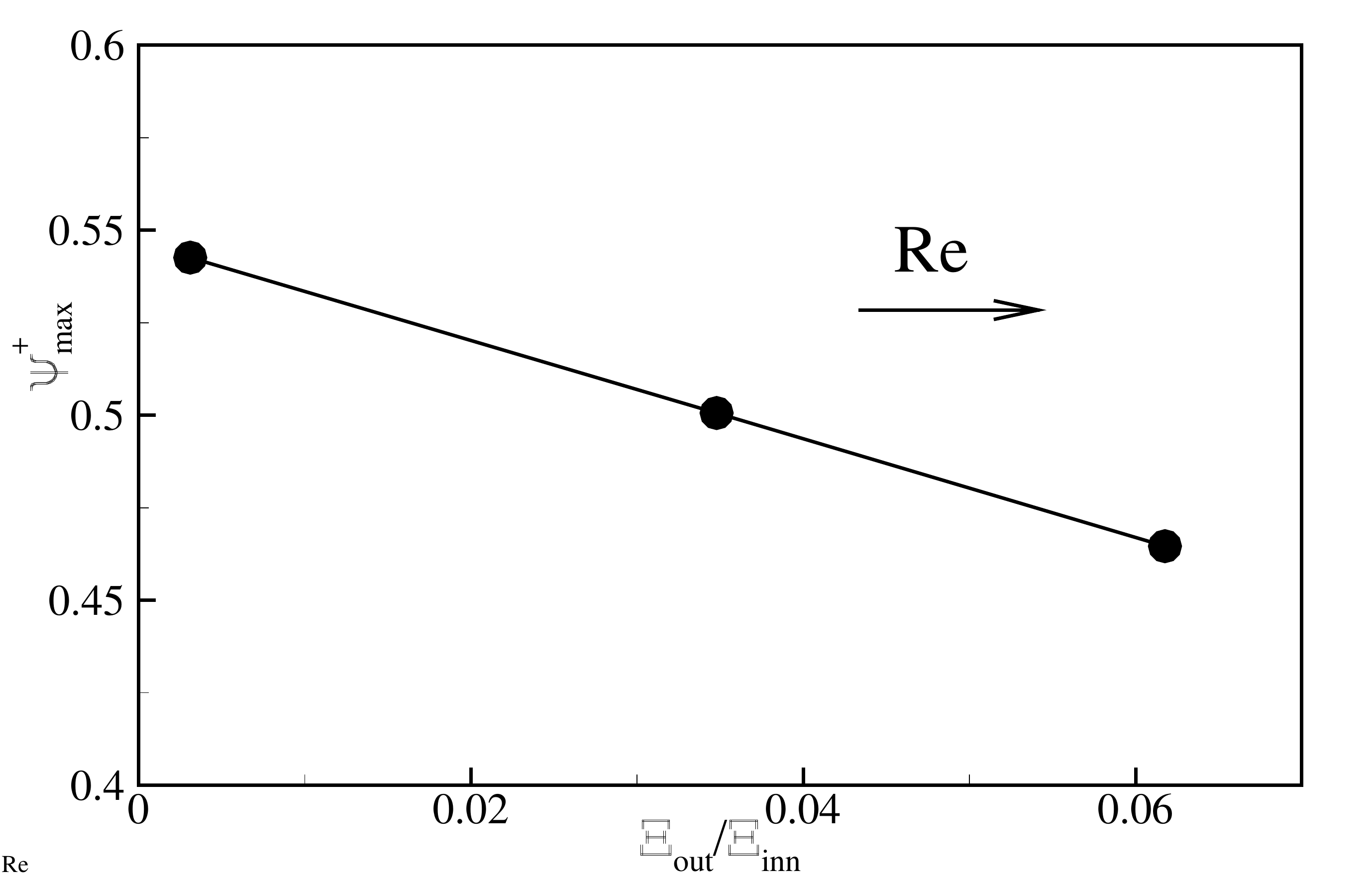}
\caption{Behavior of the near-wall peak of the single point spatial flux, $\psi_{max}^+$, equation (\ref{TKE_channel}), as a 
function of the ratio $\Xi_{out}/\Xi_{inn}$ for increasing Reynolds number.}
\label{ODSRscaling}
\end{center}
\end{figure}

In this context, the existence of a simple near-wall viscous scaling may be questioned 
by the fact that different turbulent engines with different characteristic scales are at work thus leading to anomalous scaling.
It is generally thought that the mixed inner/outer scaling is due to the fact that with increasing Reynolds number the large-scale 
structures of the overlap layer become more energetic and able to actively modulate the near-wall dynamics 
through production of near-wall fluctuations \citep{Marusic,Mathis,Marusic2010,Marusic_SCIENCE}. Complementary to this picture, 
\citet{JimenezARF} describes this modulation as a local effect where small-scales structures equilibrate with their large-scale 
environment. From a point of view of scale-energy 
source and transfer, the mixed inner/outer scaling of the near-wall region could be interpreted as the result of a 
confinement of the scale-energy excess emerging from the near-wall region due to the presence of increasingly large production scales in the 
overlap layer. At given Reynolds number, the scale-energy flux originated in the DSR and directed toward the bulk of the flow encounters the additional energy source given by the ODSR. This additional energy source radiates scale-energy and contributes to the overall energy flux. Below the ODSR the partial flux it generates is directed towards the wall, thereby opposing the  flux produced in the DSR. The result is a net decrease in the flux towards the bulk. In fact, as stated by the turbulent kinetic energy and Kolmogorov equations (\ref{TKE_channel}) and (\ref{conservation_form}), the source regions are repulsor for the fluxes. Increasing the Reynolds number, the DSR remains fixed when scaled in inner units, while the effect of the ODSR increases. As shown in figure \ref{rx0_behaviors}, the fluxes deviate to try to avoid the ODSR which involves increasingly large scales with $Re$. Hence, the overall effect is a decrease with Reynolds number of the scale-energy flux from the near wall region due to the presence of increasingly large production scales with $Re$ in the ODSR. Accordingly, in the inset of figure \ref{tke}a a decrease of the near-wall peak of the single-point spatial flux is observed at increasing $Re$. This decrease is compared with the ratio between the overall outer and inner energy source, in figure \ref{ODSRscaling}. By increasing the Reynolds number, the ratio $\Xi_{out}/\Xi_{inn}$ increases and, as a consequence, the near-wall peak of the single-point spatial flux, $\psi_{max}^+$, significanlty decreases. In conclusion, the scale-energy produced within the DSR increasingly feeds the turbulence in the near-wall region since the energy flux towards the channel center is decreasing with $Re$. The resulting growth of the energy available near the wall is, thus, responsible for intenser fluctuations with $Re$ leading to a mixed inner/outer 
scaling of near-wall quantities.

\section{Overlap layer scalings}
\label{scalings}

\begin{figure}
\begin{center}
\includegraphics[width=0.45\linewidth]{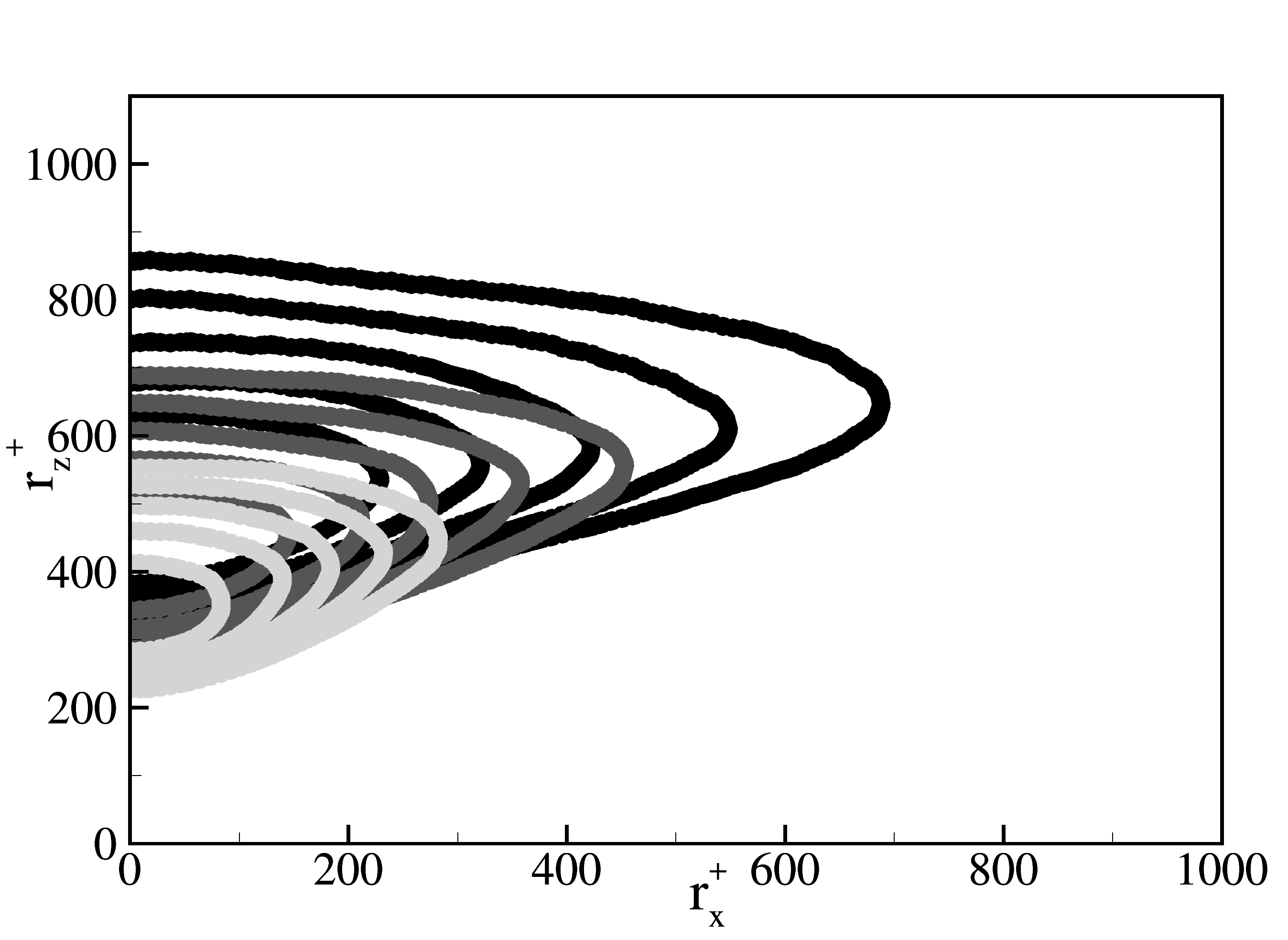}%
\mylab{-12mm}{8mm}{(\aaa)}%
\includegraphics[width=0.45\linewidth]{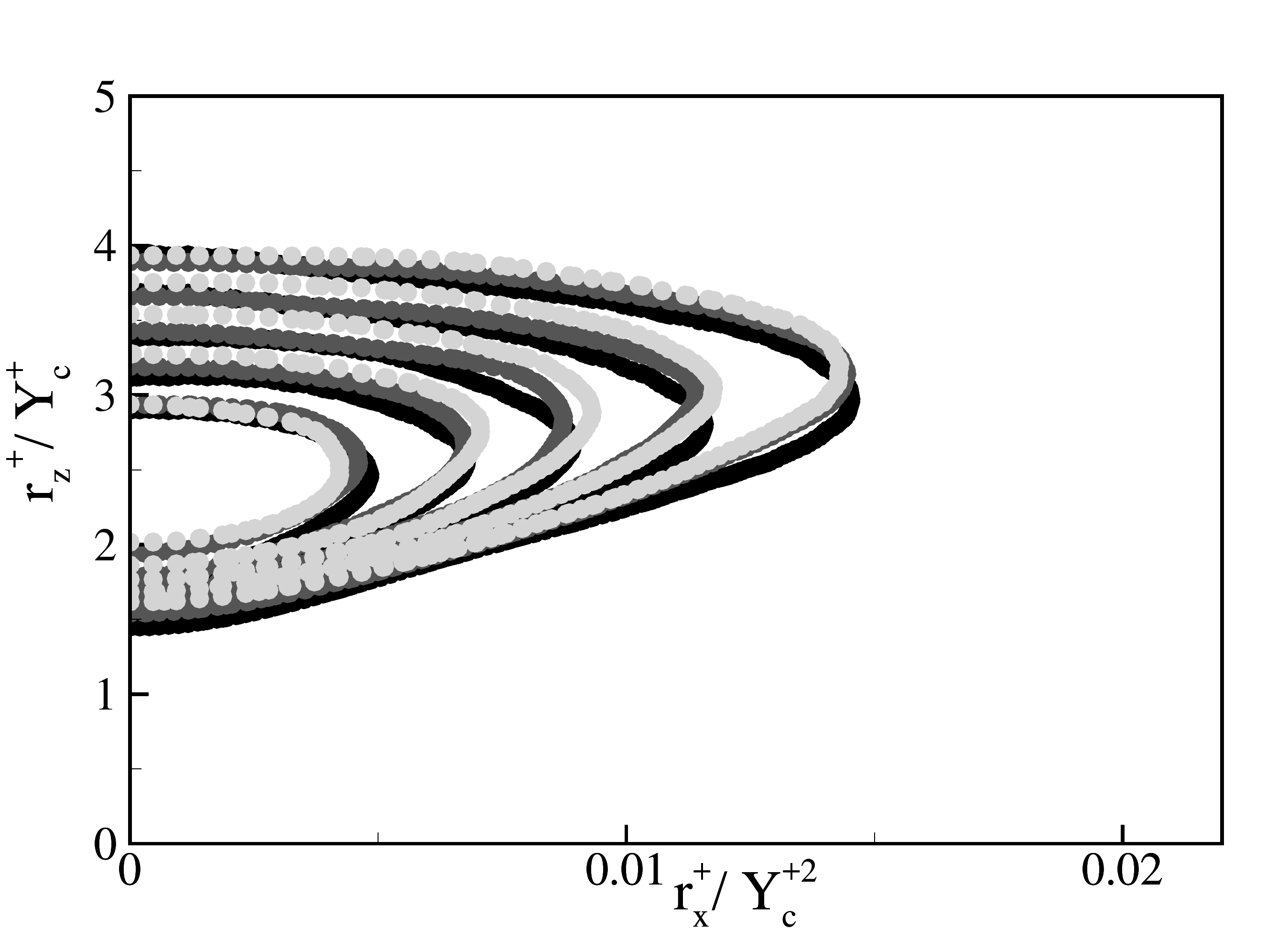}%
\mylab{-12mm}{8mm}{(\bbb)}%
\caption{Scale-energy source isolines for three distances from the wall within the overlap layer for
$Re_\tau=1500$ ($Y_c^+ = 140$ light grey, $Y_c^+ = 180$ dark grey and $Y_c^+ =220$, black). (a) In inner units 
the dimensions of the source region increase with the distance from the wall.
In (b) spanwise and streamwise scales are normalized with the wall-normal distance and its square, respectively. Apparently, the 
isolines of the scale-energy source at different wall-normal distances tend to collapse one on top of the other.}
\label{Scalings_rx_rz}
\end{center}
\end{figure}

Let us now investigate more in detail the peculiar features of the Kolmogorov equation within the overlap layer.
The first point we address is the behavior with wall distance of the spanwise scale of maximum scale-energy source for 
a given wall-distance, $\ell_z^{\xi_{max}}$. The present data show that for the three Reynolds number considered, 
$\ell_z^{\xi_{max}}$ increases almost linearly with $Y_c$, see figure \ref{scaling}(a). In particular, 
we observe $ (\ell_z^{\xi_{max}})/h \approx 0.14 + (Y_c/h)$, $(\ell_z^{\xi_{max}})/h \approx 0.14 +1.16 (Y_c/h)$ 
and $(\ell_z^{\xi_{max}})/h \approx 0.14 + 1.4 (Y_c/h)$ from low to high Reynolds numbers respectively. 
This behavior is similar to that reported in \citet{Neela}
for the shear scale $L_s$ which is found to slightly increase its slope from $Re_\tau=300$ up to $Re_\tau=2000$ where it seems to asymptotically approach 
the dimensional prediction $L_s=ky$. Contrary to the slope, the intercept remains almost constant. This value of the intercept could be considered as the 
characteristic spanwise scale of the lower overlap layer. As shown by the isolines in figure \ref{scaling}(b) for 
$Re_\tau=1500$, in the overlap layer the iso-levels of positive scale-energy source, $\xi>0$, intercept 
spanwise scales which linearly increase with the distance from the wall. In particular, the iso-levels 
of $\xi$ can be approximated by a sheaf of lines originating from a unique point at $\tilde{r}_z^+ = 120$ and $\tilde{Y}_c^+ = 80$ but 
with different slopes, i.e. $r_z^+ = \gamma (\xi_0^+) (Y_c^+ - \tilde{Y}_c^+) + \tilde{r}_z^+$ where $\xi_0^+$ defines the iso-level of $\xi$
and $\gamma(\xi_0^+) \in [0.04 ; 1.4]$.

As already stated, the source maximum is found at $r_x=0$ for all $Y_c$ while for $r_x> 0$ the source decreases, see 
figure \ref{Scalings_rx_rz}(a). By tracking the spanwise scale of the maximum of $\xi$ as a function of the streamwise 
scale, we find that these spanwise scales increase following a square-root law, $r_z^+ \sim \sqrt{r_x^+}$, independently of 
the Reynolds number. This behavior is similar to that reported by \citet{Jimenez_spectra} for the spectral 
distribution of the Reynolds stresses and it finds a possible theoretical explanantion in \citet{moarref2013} in terms of 
geometrically self-similar resolvent modes. Since we observe that the spanwise scales involved in the scale-energy source 
increase linearly with wall-distance, $r_z^+ \sim Y_c^+$, we argue that the streamwise scales should behave quadratically 
with wall-distance, i.e. $r_x^+ \sim Y_c^{+2}$. Hence, we expect 
the scale-space behavior of the scale-energy source of the ODSR in the overlap layer to be approximatively self-similar
if plotted as function of $r_x^+/Y_c^{+2}$ and $r_z^+/Y_c^+$. As shown in figure \ref{Scalings_rx_rz}(b) this rescaling of $\xi$ 
allows us for a unique comprhensive view of the outer scale-energy source where the $Y_c$-dependence is dropped. The data we have 
available at other Reynolds numbers (not shown) suggest that this behaviour is $Re_\tau$ independent. The 
comparison with figure \ref{Scalings_rx_rz}(a) highlights that this single picture of the overall behavior of 
the outer scale-energy source would be missed when using viscous units. 
It is worth mentioning that the observed scale-space distribution of the outer scale-energy source, $r_z^+ \sim Y_c^+$ and $r_x^+ \sim Y_c^{+2}$, is consistent with the energy distribution found in \citet{moarref2013} which, although different from the original scaling proposed by \citet{Townsend}, is explained by the conjecture of an overlap layer populated by self-similar structures attached to the wall, see e.g. \citet{delAlamo}.

\begin{figure}
\begin{center}
\includegraphics[width=0.45\linewidth]{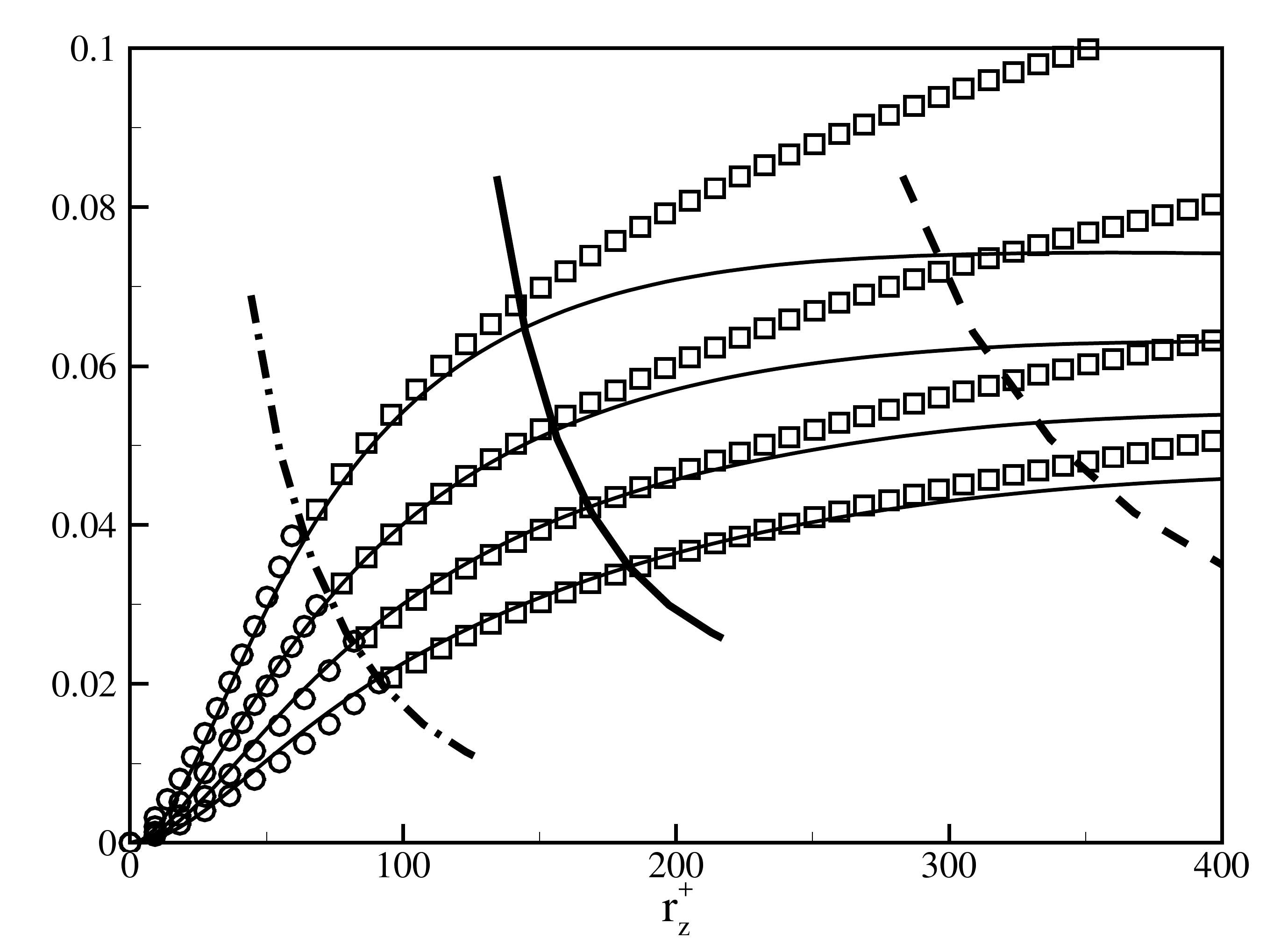}%
\mylab{-52mm}{36mm}{(\aaa)}%
\includegraphics[width=0.45\linewidth]{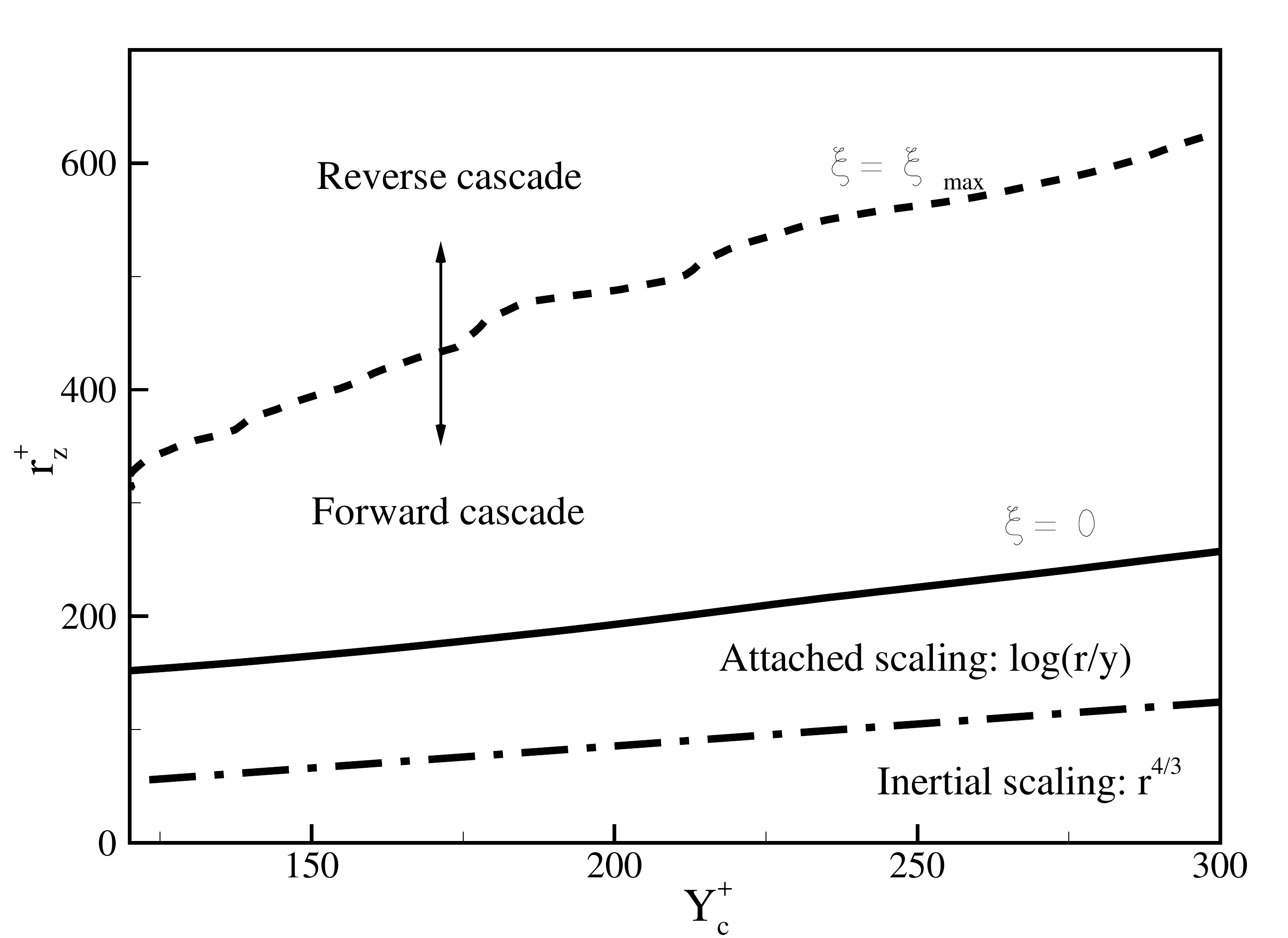}%
\mylab{-52mm}{36mm}{(\bbb)}%
\caption{(a) Scaling of production, $-2 \langle \delta u \delta v \rangle (dU / dy)$, for $r_x=0$ and $Re_\tau=1500$. From top to bottom the four solid lines provide the production at $Y_c^+ = 130$, $160$, $190$ and $230$, respectively. Squares denote the logarithmic 
behavior described by equation (\ref{Prodlog}) and circles denote the inertial behavior, equation (\ref{Prod}). The three thick 
lines show the characteristic scales reported in (b). (b) $Y_c$-behavior of the scale of maximum 
scale-energy source, $(\ell_z^{\xi_{max}})^+ \approx 215 + 1.4 Y_c^+$ (dashed line) and of the zero scale-energy source, 
$(\ell_z^{\xi_{zero}})^+ \approx 80+ 0.65 Y_c^+$ (solid line). The shear scale,
$\ell_S^+ = \kappa Y_c^+$ (dashed-dotted line), represents the cross-over 
between logarithmic and power law scaling of production.}
\label{Scalings_log}
\end{center}
\end{figure}

We address now the possibility to scale also the intensity of the outer scale-energy source, $\xi = -2 \langle \delta u \delta v \rangle \left ( dU / dy \right )^* - 4\langle \epsilon^* \rangle$. The rate of viscous dissipation is constant in the space of scales and can be easily modeled by means of the equilibrium hypothesis, $\langle \epsilon \rangle \sim u_\tau^3 / \kappa y$. On the contrary, the scale-dependent behaviour of the production intensity is not trivial and need a detailed analysis. 
Two distinct regimes are expected whose transition should be controlled by the shear scale $\ell_S$ \citep{Jacob}. For scales larger than $\ell_S$ the prevailing mechanism which determines the scaling law is production. 
For scales smaller than $\ell_S$, the energy cascade prevails and an isotropy-recovering behavior is expected to occur. This range should be characterized by a power law with universal exponents 
based on the dimensional predictions proposed by \citet{Lumley}. As shown in figure \ref{Scalings_log}(a) for $Re_\tau = 1500$ 
(the same behavior is observed also for the lower Reynolds numbers considered), the scale-energy production 
follows the classical Lumley's prediction for the mixed structure 
function \citep{Lumley,Jacob}, here extended to scale-energy production by taking into account 
the mean shear, $dU /dy$, 
\begin{equation}
-2 \langle \delta u \delta v \rangle \frac{dU}{dy}= \beta u_\tau^3 (r/y)^{4/3}/\kappa y 
\qquad \mbox{for} \qquad r < \kappa y \, ,
\label{Prod}
\end{equation}
where $\kappa$ is the von K\'arm\'an constant, see circles in figure \ref{Scalings_log}(a). Production follows this power law for scales smaller than the shear scale 
$\ell_S$ as shown by the thick dashed-dotted line in figure \ref{Scalings_log}(a). The shear scale defined as $\ell_S = \sqrt{\epsilon/S^3}$ where $S= dU/dy$, is computed here by using the overlap layer estimate, $\ell_S = \kappa Y_c$. Hence, the power law (\ref{Prod}) is valid for 
scales smaller than the distance from the wall, the so-called {\it detached} scales. At scales larger than $\ell_S$ ({\it attached} scales), we find that the production 
in the overlap layer is well described by a logarithmic law,
\begin{equation}
-2 \langle \delta u \delta v \rangle \frac{dU}{dy}= u_\tau^3 (C+D \log(r/y))/\kappa y 
\qquad \mbox{for} \qquad \kappa y < r < \ell_z^{\xi_{zero}} \, ,
\label{Prodlog}
\end{equation}
squares in figure \ref{Scalings_log}(a).
This law is closely related to the ($k^{-1}$)-law for the energy spectrum derived by \citet{Perry2}, see also \citet{Nikora}. As shown in \citet{Davidson}, the
real space analogue of the ($k^{-1}$)-law is a logarithmic law for the streamwise second-order structure functions. 
Equation (\ref{Prodlog}) represents an extention of this law to the mixed structure functions, 
$\langle \delta u \delta v\rangle$, which takes into account also the mean shear, $dU / dy$. Scaling (\ref{Prodlog}) remains valid  
up to $r_z = \ell_z^{\xi_{zero}}$ where $\ell_z^{\xi_{zero}}$ is defined as the scale where the source term becomes zero, 
$\xi=0$. For scales smaller than $\ell_z^{\xi_{zero}}$ the source term is actually a sink for the scale-energy fluxes, 
$\xi < 0$, while for larger scales it is a source, $\xi>0$.

\begin{figure}
\begin{center}
\includegraphics[width=0.3\linewidth]{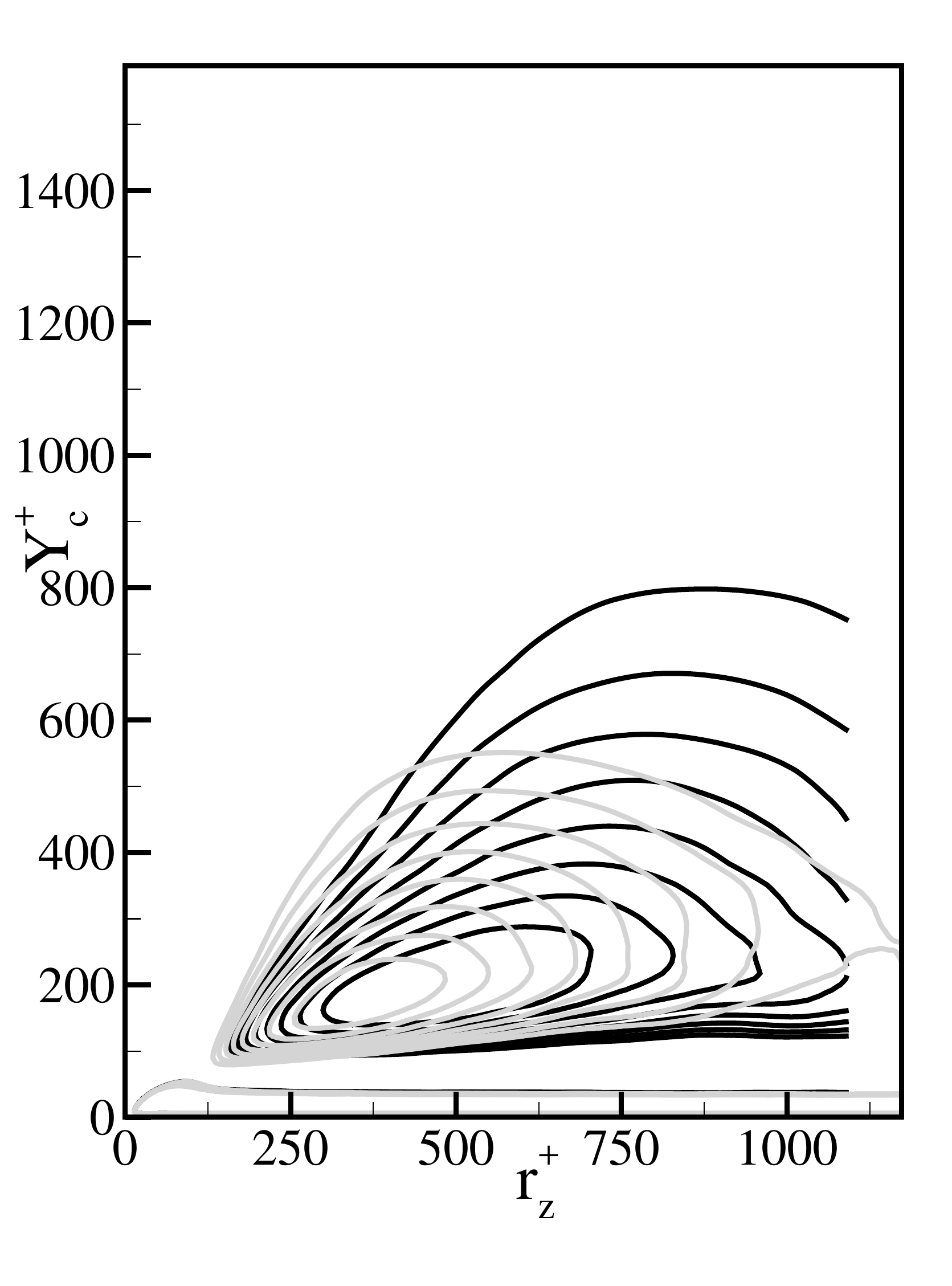}%
\mylab{-33mm}{45mm}{(\aaa)}%
\includegraphics[width=0.3\linewidth]{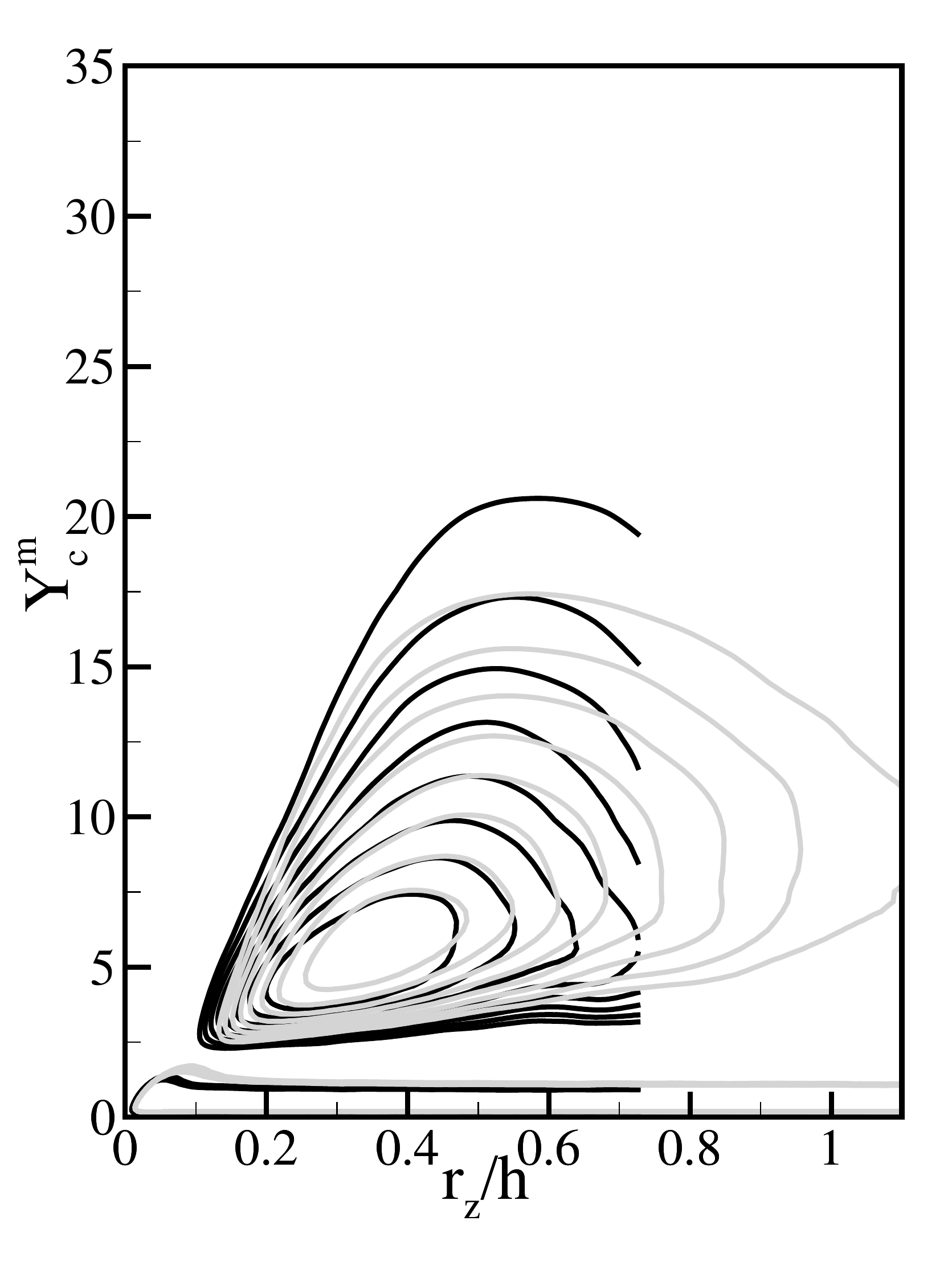}%
\mylab{-33mm}{45mm}{(\bbb)}%
\includegraphics[width=0.3\linewidth]{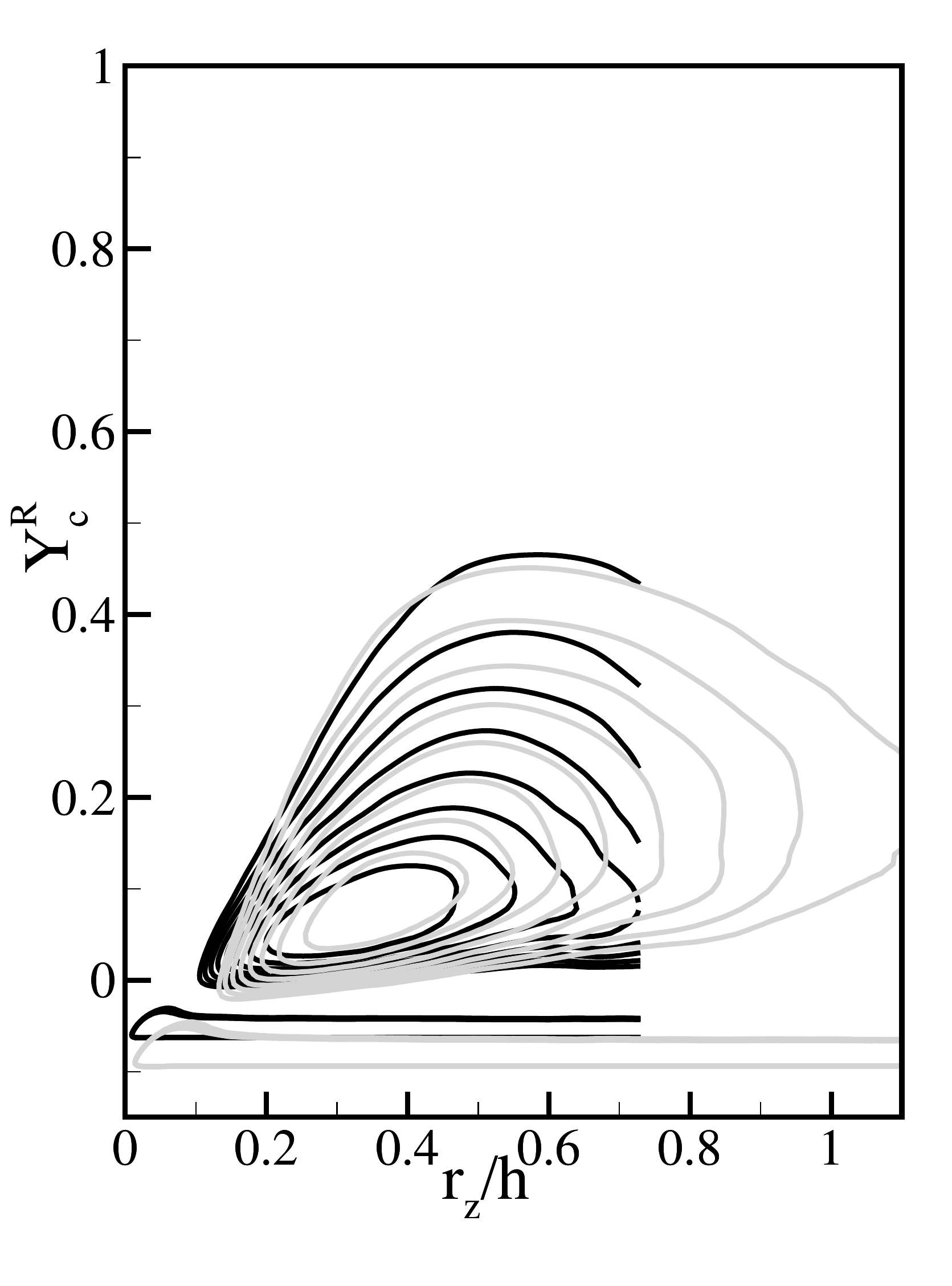}%
\mylab{-33mm}{45mm}{(\ccc)}%
\caption{Energy source -- $\xi^+$  -- isolines in the $r_x=0$ plane for $Re_\tau=1000$ (grey) and
$Re_\tau=1500$ (black). In (a) $\xi^+$ is shown as a function of viscous quantity,
$(r_z^+, Y_c^+)$, in (b) as a function of mixed quantity, $(r_z/h, Y_c^m)$, and in (c)
as a function of rescaled quantity, $(r_z/h, Y_c^R)$.}
\label{Scalings_OSR}
\end{center}
\end{figure}

According to equation (\ref{conservation_form}) and as shown by the vector field in 
figure \ref{rx0_behaviors}, scales smaller than $\ell_z^{\xi_{zero}}$ are characterized by a negative divergence of the 
fluxes since $\xi < 0$ and, thus, by a forward cascade. 
As a consequence, the logarithmic behavior (\ref{Prodlog}) is associated with turbulent scales involved in a 
forward cascade process. Since, as anticipated, these scales are larger than $\ell_S = \kappa Y_c$ 
they are influenced by the presence of the walls.
On the other hand, the power law scaling 
characterizes scales smaller than $\ell_S = \kappa Y_c$. Hence, the influence of the wall should be somehow negligible and,
consequenlty, the forward cascade should resemble the classical one proposed by Kolmogorov since an isotropic recovery 
should start to take place, see \citet{Casciola_2005}. It could be worth stressing that the shear scale, often identified with the distance from the wall, has been longly addressed to explain the momentum and energy transfer between different scales of motion, starting from the pioneering work of \citet{Townsend} to more recent contributions, see e.g. \citet{Nikora} and \citet{LozanoDuranJFM2014}. As shown in figure \ref{Scalings_log}(b), both $\ell_S$ 
and $\ell_z^{\xi_{zero}}$ linearly increase with the wall distance denoting again statistical features attached to the wall. 
Since the slope of $\ell_z^{\xi_{zero}}$ is larger than that of the shear scale, $\ell_S$, we can expect that the range of 
scales where the logarithmic scaling holds, should expand significantly with Reynolds number increasing the importance of 
the scaling for the prediction of production.

Figure \ref{scaling} suggests a possible approximation for the source term for scales larger than $\ell_z^{\xi_{zero}}$. By assuming that the production term for scales close to $\ell_z^{\xi_{max}}$ roughly behaves 
linearly with $r_z$, see figure \ref{Scalings_log}, we can model the source term as $\xi^+ \approx A(Y_c^+) r_z^+ + B(Y_c^+)$ since 
dissipation is a scale-independent process. In addition, we have also found in section \ref{OSR} that the iso-levels of $\xi > 0$ form a 
sheaf of lines described by the equation, $r_z^+ = \gamma (\xi_0^+) (Y_c^+ - \tilde{Y}_c^+) + \tilde{r}_z^+$ where $\xi_0^+$ defines the iso-level of $\xi$ we are considering, see figure \ref{scaling}. By combining these two informations, the slope $\gamma$ is given by 
$\gamma (\xi_0^+) \approx G \xi_0^+ + H$, implying that the source term can be approximated as 
$\xi^+ \approx (r_z^+ - \tilde{r}_z^+)/ G (Y_c^+ - \tilde{Y}_c^+) - H/G$. From our data we find $G=103.9$ and $H=0.58$.

Summarizing, we have shown that for a fixed Reynolds number the complex multidimensional features of the ODSR can be described 
in a simple way by using the self-similarity of the streamwise and spanwise scales shown in figure \ref{Scalings_rx_rz} and the 
scaling of the production intensity shown in figure \ref{Scalings_log}. Despite the limited range of Reynolds numbers we 
have available, let us address now how this picture changes for different Reynolds numbers. As already 
shown in section \ref{OSR}, the intensity of the outer scale-energy source scaled in viscous units and its scale-space location 
scaled in outer units remain constant for the present Reynolds numbers. Hence, the effect of Reynolds number mainly reduces to an 
expansion/shrinking of the wall-normal distances. 
A sufficiently general procedure could be to consider a mixed scaling of the wall-normal  
distance in the form $Y_c^m= ({Y_c}/h)^\alpha ({Y_c^+})^\beta$. Alternatively one could use outer units  
shifting the origin according to the expression $Y_c^R = (Y_c/h) - \eta^+ / Re_\tau$ where $\eta^+$ is a free parameter.
By considering $\alpha = \beta = 1/2$ in the first case and $\eta^+= 100$ in the second one,
the resulting scaling of the outer scale-energy source is shown in figure \ref{Scalings_OSR}. Although not conclusive given the small range of 
Reynolds numbers investigated this result seems to be consistent with the
possibility of a universal rescaling of the outer scale-energy source (ODSR).

\section{Concluding remarks}

By means of the description of turbulence given by the generalized Kolmogorov equation we study the 
scale-energy transfer and production mechanisms of turbulent wall flows at different Reynolds numbers.
Two driving mechanisms in terms of scale-energy source are identified for the fluxes. The first stronger one, the 
driving-scale range DSR, belongs to the near-wall cycle. As expected, its inner-scaled topology remains unaltered 
with Reynolds while its intensity is found to slightly increase with $Re$ (near wall modulation). The second outer 
scale-energy source, outer driving scale range ODSR, takes place further away from the wall in the overlap layer and 
is separated from the DSR by a distinct scale-energy sink layer suggesting a possible independence of the 
production mechanisms of the ODSR from the near-wall region which might be interpreted as an autonomous outer cycle. 
Although its intensity is small compared to the DSR, the outer region of scale-energy source expands with Reynolds number
while its peak intensity remains almost constant. 
These observations suggest the importance of the ODSR for large Reynolds number wall-turbulence.

Further analysis of the ODSR demonstrates that the $Re$-dependence of the outer scale-energy source can be dropped 
by scaling in outer units the space of scales and in mixed one the wall-distance at least for the range of Reynolds numbers 
analysed here.
Furthermore, we found that the spanwise scales involved in the scale-energy source linearly increase with the 
distance from the wall. On the other hand, the streamwise scales are connected to these spanwise scales of scale-energy source 
through a square root law and, hence, quadratically increase with wall distance. These observations 
allow us to scale the outer scale-energy source highlighting its self-similarities for different wall distances and 
Reynolds numbers. 
While considering the intensity of the outer scale-energy source, we found that 
the space of scales within the overlap layer can be divided into two distinct ranges. For scales larger than the 
shear scale, $\ell_s$, but smaller than the cross-over scale of zero scale-energy 
source, $\ell_z^{\xi_{zero}}$, the outer scale-energy source follows a logarithmic law, $\xi = u_\tau^3 (C+ D \log(r/y) -1)/\kappa y$. This behavior is theoretically consistent with the presence of a $k^{-1}$ law for the energy cospectrum.
For scales smaller than the shear scale, the outer scale-energy source follows a power law whose exponent equals 
Lumley's dimensional prediction, $\xi = u_\tau^3 (\beta (r/y)^{4/3} -1)/\kappa y$. These scales are involved in a direct cascade whose features should 
resemble the classical one since they are detached from the wall and, hence, an isotropic recovery is expected to take place. 
Interestingly, both $\ell_z^{\xi_{zero}}$ and $\ell_s$ increase linearly with wall distance. The different 
increase with wall-distance of these two scales highlights the possible extention of the range of scales of validity of 
the logarithmic law for the prediction of outer scale-energy source at large Reynolds numbers. Overall, these observations 
suggest a strong connection of the observed outer scale-energy source with the presence of an outer turbulence production 
cycle whose statistical features agree with the hypothesis of an overlap layer dominated by self-similar structures attached to the wall.

The topology of the energy transfer is also studied. 
The paths of energy resembles the one reported in \citet{Cimarelli_JFM2} for a lower Reynolds number case. Only 
one singularity point related to the DSR exists from which the fluxes depart also for larger Reynolds number. 
According to our observations we may expect a high Reynolds number state of wall-turbulence where only one  
origin for the fluxes exists and corresponds to the DSR at the small-scales of the near-wall region, since the 
intensity of the ODSR should be substantially $Re$-independent. 
In this scenario, the Reynolds number effects on the energy transfer should come only from the expansion 
of the ODSR both in scale and physical space with $Re$. For increasing Reynolds numbers, the turbulent energy emerging from the 
DSR near the wall experiences an expanding outer scale-energy source in the overlap layer which acts as a 
repulsor for the fluxes as stated by equation (\ref{conservation_form}). Hence, the fluxes try to avoid the 
increasingly large production scales of the ODSR and remain partially confined to the wall-region increasing the overall amount of energy locally 
available near the wall. Accordingly, we observe a decrease of the spatial flux from the buffer to the overlap layer 
at increasing Reynolds numbers. Hence we may conjecture that the near-wall modulation is a result of a confinement 
of the near-wall source due to the presence of increasingly large production scales in the overlap layer.

\bibliographystyle{plain}
\bibliography{JFM_high_re}

\end{document}